\documentclass[11pt]{article}
\pdfoutput=1
\usepackage{jheppub}
\usepackage[table]{xcolor}
\usepackage{graphicx, psfrag}
\usepackage{dynkin-diagrams}
\usepackage{appendix}
\usepackage{hyperref}
\usepackage{color}
\usepackage{booktabs,caption}
\usepackage[flushleft]{threeparttable}
\usepackage{lscape}
\definecolor{Mygrey}{gray}{0.8}
\definecolor{Mywhite}{gray}{1.0}

\newcommand{\be}{\begin{equation}}
\newcommand{\ee}{\end{equation}}
\newcommand{\bea}{\begin{eqnarray}}
\newcommand{\eea}{\end{eqnarray}}
\title{Classifying three-character RCFTs with Wronskian Index equalling $\mathbf{0}$ or $\mathbf{2}$}

\author[a]{Arpit Das,}
\author[b,c]{Chethan N. Gowdigere}
\author[b,c]{and Jagannath Santara}

\affiliation[a]{Centre for Particle Theory, Department of Mathematical Sciences,\\
Durham University, South Road, Durham DH1 3LE, UK}
\affiliation[b]{National Institute of Science Education and Research Bhubaneshwar,\\
P.O. Jatni, Khurdha, 752050, Odisha, INDIA}
\affiliation[c]{Homi Bhabha National Institute, Training School Complex,\\
Anushakti Nagar, Mumbai 400094, INDIA}

\emailAdd{arpit.das@durham.ac.uk}
\emailAdd{chethan.gowdigere@niser.ac.in}
\emailAdd{jagannath.santra@niser.ac.in}

\abstract{In the modular linear differential equation (MLDE) approach to classifying rational conformal field theories (RCFTs) both the MLDE and the RCFT are identified by a pair of non-negative integers $\textbf{[n,l]}$. $\mathbf{n}$ is the number of characters of the RCFT as well as the order of the MLDE that the characters solve and $\mathbf{l}$, the Wronskian index, is associated to the structure of the zeroes of the Wronskian of the characters. In this paper, we study  $\textbf{[3,0]}$ and $\textbf{[3,2]}$ MLDEs in order to classify the corresponding CFTs. We reduce the problem to a ``finite'' problem: to classify CFTs with central charge  $ 0 < c \leq 96$, we need to perform  $6,720$ computations for the former and $20,160$ for the latter. Each computation involves (i) first finding a simultaneous solution to a pair of Diophantine equations and (ii) computing Fourier coefficients to a high order and checking for positivity. 

In the $\textbf{[3,0]}$ case, for $ 0 < c \leq 96$, we obtain many character-like solutions: two infinite classes and a discrete set of $303$. After accounting for various categories of known solutions, including Virasoro minimal models, WZW CFTs, Franc-Mason vertex operator algebras and Gaberdiel-Hampapura-Mukhi novel coset CFTs, we seem to have seven hitherto unknown character-like solutions which could potentially give new CFTs. We also classify $\textbf{[3,2]}$ CFTs for $ 0 < c \leq 96$: each CFT in this case is obtained by adjoining a constant character to a $\textbf{[2,0]}$ CFT, whose classification was achieved by Mathur-Mukhi-Sen three decades ago. }

\begin{document}
\maketitle

\flushbottom

\section{Introduction \label{1s}}

Two dimensional conformal field theory (CFT) is a subject of primary importance  relevant to many areas in physics and mathematics; \cite{DiFrancesco:1997nk, Moore:1989vd, Fuchs:2009iz, Gaberdiel:1999mc} comprise a  partial list of references. In physics it is important for string theory both in perturbation theory, initially and later for various aspects of non-perturbative string theory also. Two dimensional CFTs describe two dimensional critical systems at  RG fixed points. They  have contributed to many areas of mathematics including representation theory, infinite dimensional algebras, theory of modular forms, etc.  An important subclass of two dimensional conformal field theories is that of  rational conformal field theories (RCFTs). In these theories the central charge $c$ and the conformal dimensions of the primary fields, the  $h$s, are all rational numbers \cite{Anderson:1987ge}. An infinite number of RCFTs are known to exist. For example the infinite series of Virasoro minimal models \cite{Belavin:1984vu} (see also \cite{DiFrancesco:1997nk}), and  the infinite set of Wess-Zumino-Witten conformal field theories (WZW CFTs) \cite{Witten:1983ar}. Even though an infinite number of RCFTs are known, there is no known  classification of them. 

One classification scheme for RCFTs is based on the structure of their torus partition functions. For a RCFT, the torus partition function can be expressed as a sum of holomorphically factorised terms. The holomorphic factors are the characters of the RCFT and the number of characters, denoted by $n$, is an important detail of a RCFT for this classification scheme. Another important detail of a RCFT relevant for this classification scheme arises as follows. The $n$ characters are the linearly independent solutions of a single $n$-th order ordinary differential equation on the moduli space of a torus, a modular linear differential equation (MLDE) \cite{Eguchi:1987qd, Mathur:1988gt, Anderson:1987ge}. One then considers the Wronskian of these solutions which is a modular function of weight $n (n-1)$. The Wronskian index of the RCFT, denoted by $l$, is a number that takes non-negative integer values and is associated with the structure of zeroes of the Wronskian and can be expressed purely in terms of the RCFT data (the central charge, the number of characters and the conformal dimensions of the primary fields). In this classification scheme, a RCFT is identified in terms of these two numbers $\mathbf{[n, l]}$, the number of characters and the Wronskian index. 

Remarkably, and this is the primary reason this classification scheme exists, the form of the MLDE gets fixed by this same pair $\mathbf{[n, l]}$. Actually  the coefficient functions of the MLDE are fixed by $\mathbf{[n, l]}$, apart from some undetermined parameters. Thus two RCFTs with the same $\mathbf{[n, l]}$ values are both solutions to the same MLDE but for a different set of parameters in the MLDE.  One thus sets up a MLDE for a given number of characters and a given Wronskian index and studies all character-like solutions (that is solutions with non-negative integer coefficients in their $q$-expansion) and obtains all RCFTs with that number of characters and that Wronskian index and thus classifies them. This is the MLDE approach to the classification of RCFTs, first carried out in \cite{Mathur:1988na}, where a  $\mathbf{[2, 0]}$ MLDE was studied and all $\mathbf{[2, 0]}$ RCFTs were classified.  The program has been implemented for $\mathbf{[2, 2]}$ first in \cite{Naculich:1988xv} and then in \cite{Hampapura:2015cea, Gaberdiel:2016zke, Hampapura:2016mmz}; then partially for three-character RCFTs in \cite{Mathur:1988gt, Mukhi:2020gnj, Tener:2016lcn, kaneko4, CM:2011, PSEZ:2016}. The MLDE approach to RCFTs has been covered in both the physics \cite{Hampapura:2015cea, Gaberdiel:2016zke, Mukhi:2020sxt, Bantay:2010uy, Tener:2016lcn, Harvey:2018rdc, Harvey:2019qzs, Bae:2017kcl, Bae:2020xzl, Kaidi:2020ecu, Kaidi:2021ent, Bae:2021mej}  and the maths literature \cite{kaneko5, kaneko1, kaneko2, kaneko3, Gannon:2013jua, arike1, franc1, kaneko4, mason1}. A status report on the program can be found in \cite{Mukhi:2019xjy}. An extension of this program that includes fermionic RCFTs and up to three characters can be found in \cite{Bae:2020xzl, Bae:2021mej}.

In this paper, we take on the task of classifying three-character CFTs. We do restrict ourselves to the cases of $l = 0$ and $l = 2$, i.e. $\mathbf{[3, 0]}$ and $\mathbf{[3, 2]}$ CFTs. The study of three-character CFTs are at the least one level harder technically than two-character CFTs. One aspect of this can be seen from the form of the MLDEs involved. In a MLDE \eqref{7n}, the coefficient functions $\phi_r$ which are modular w.r.t the full modular group are fixed by the Wronskian indiex $l$; the functional form is completely fixed leaving only a few arbitrary parameters. For two character MLDEs studied so far in the literature, viz. $\mathbf{[2, 0]}$, $\mathbf{[2, 2]}$ and $\mathbf{[2, 4]}$, the  MLDE has only one arbitrary parameter (the $\mathbf{[2, 4]}$ MLDE does have two parameters to begin with but one of them gets fixed by other considerations \cite{Chandra:2018pjq}). On the other hand both the $\mathbf{[3, 0]}$ and $\mathbf{[3, 2]}$ MLDEs have two arbitrary parameters. Solving the MLDE means also finding the value of the parameters for which solutions exist. Hence obtaining two-character RCFTs  involves scanning the set of possibilities for one parameter which is the infinite set of all rational numbers. While solving for three-character RCFTs (for $\mathbf{l} \leq 2$), one would need to scan over the infinite set of \emph{pairs} of rational numbers.

There is another sense in which the problem of classifying three-character CFTs is harder. There already exists  an infinite number of such  CFTs. In a recent work \cite{Das:2020wsi}, with the intention of correctly placing the known classes of CFTs within the present classification scheme, we computed the $\mathbf{[n, l]}$ values  for an infinite number of CFTs. This allows us to obtain a partial classification of all CFTs with a given $\mathbf{[n, l]}$. We call this a partial classification because it partially answers the question ``what is the classification for all CFTs with this given $\mathbf{[n, l]}$?'' The partial classification for $\mathbf{[3, 0]}$ CFTs from that paper is given here \eqref{63s}. As one can see, there are two infinite classes (second and fourth lines), $7$ irreducible CFTs and $8$ tensor product CFTs. Thus, even before we start the problem, we are told that there will be an infinite number of solutions, at least. This should be contrasted with what happens for $\mathbf{[2, 0]}$ CFTs. The partial classification from \cite{Das:2020wsi} reveals that there are a finite number of such CFTs viz. $8$ (and it eventually turns out that the full classification \cite{Mathur:1988na} also has only these $8$ CFTs and nothing else). Here in this  paper,  at the end of the day, we do reproduce the partial classification of \cite{Das:2020wsi} for $\mathbf{[3, 0]}$ CFTs  and find two other infinite classes of solutions as well as a discrete set of CFTs and we may have a few ($7$) potentially new CFTs. The subsequent case of $\mathbf{[3, 2]}$ CFTs is intriguing in it's own way. There is not a single such CFT among the infinite number of CFTs that we studied in \cite{Das:2020wsi}.  Here in this paper, we do manage to classify all  $\mathbf{[3, 2]}$  CFTs up to central charge $96$ and we find exactly nine new CFTs. But in a sense, they are not as interesting since it seems that they are related to the nine  $\mathbf{[2, 0]}$ admissible character solutions and it is not clear if they should be considered as new CFTs.  In any case, we have achieved classification.  

In this paper, we obtain solutions to the MLDEs by directly solving them; the MLDE is a homogenous linear differential equation and one can solve it with the Frobenius method. Although we do give a way of thinking about the MLDE as an inhomogenous MLDE (one order lower), which turns out to be useful to  understand features of some of the solutions, our main method is the direct Frobenius one.  One directly plugs in the $q$-expansion of the characters: $ \chi_i(q)= q^{\alpha_i}\sum\limits_{n=0}^{\infty} f^{(i)}_n q^n, ~f^{(i)}_0 \neq 0$ into the MLDE and solve it order by order. At the lowest order, one gets a $\mathbf{n}$th order polynomial equation, the indicial equation, whose roots are the indices $\alpha_i$. Higher orders give  expressions for the $f^{(i)}_n$'s which are rational functions of the parameters of the MLDE and the index $\alpha_i$. Then one focusses on the identity character whose index $\alpha_0$ (for unitary conformal field theories) is related to the central charge  ($\alpha_0 = - \frac{c}{24}$) and each of it's Fourier coefficients are (non-negative) integers. There is a lot of purchase that comes out of this integrality. Typically one imposes the fact that $m_n$'s (the coefficients of $q^n$ in the identity character) are integers (just for a few $n$'s) which constrains the possibilities for $\alpha_0$. One kind of equation that comes out of this analysis is a polynomial equation for $\alpha_0$ with integral coefficients;  the integral-root theorem then tells us the the kind of rational values that $\alpha_0$ and hence $c$ can take. For example in the $\mathbf{[2, 0]}$ and the $\mathbf{[2, 2]}$ cases \cite{Hampapura:2015cea},  the integrality of  $m_1$ implies that for $c = \frac{p}{q}$ with $p$ and $q$ coprime, $q$ can be either $1$ or $5$ (we give more details for $\mathbf{[2, 0]}$ in section \ref{22ss} below).  This means that $5c$ is always an integer; hence for a fixed range of central charge, say $0 < c \leq 1$, there are now only $5$ possibilities viz, $\frac15, \frac25 \ldots \frac55$. Prior to this stage of the analysis, we had an infinite number of possibilities for $c$, even for a finite range. Hence, the problem from here on, we say, has become a ``finite'' problem.   For the $\mathbf{[3, 0]}$ and the $\mathbf{[3, 2]}$ cases, we have to impose integrality of both $m_1$ and $m_2$ to get a polynomial equation for $\alpha_0$ with integral coefficients. It then turns out that, for the $\mathbf{[3, 0]}$ case, $70c$ is an integer, the problem becomes ``finite''  and for $0 < c \leq 1$ 
there are $70$ possibilities $\frac{1}{70}, \frac{2}{70} \ldots \frac{69}{70}, \frac{70}{70}$. In this paper, we study CFTs with $c \leq 96$ and hence consider $96 \times 70 = 6720$ possibilities. For the $\mathbf{[3, 2]}$ case, the corresponding number is $96 \times 210 = 20160$\footnote{It turns out that for the $\mathbf{[2,0]}$ case there is a further constraint for the central charge viz. $c < 10$, which makes even the range of the central charge finite.  This makes the number of possibilities actually finite: $49$. There is a similar constraint in the $\mathbf{[2,2]}$ case \cite{Hampapura:2015cea} that makes the range of the central charge itself finite and again results in a finite number of possibilities for $c$. For three-character CFTs at least for the $\mathbf{[3,0]}$ case, we cannot expect such a finiteness for the range of $c$ simply because even the known CFTs in this class have an infinite range of central charge \eqref{63s}.}. This strategy of making the problem ``finite'' is a very  important first step in our procedure to classify three-character CFTs. 
Further steps are described in the subsequent sections; they involve, for each of these finite possibilities, finding simultaneous solutions to a pair of Diophantine equations, computing higher order Fourier coefficients and checking for positivity. 

\emph{Disclaimer} : In this paper, when we say, we solved a certain MLDE and obtained a CFT, what we really would have done is, we solved that MLDE and obtained a character-like solution a.k.a admissible character solution i.e. a solution whose Fourier coefficients are non-negative. Strictly speaking,, we should say we solved the MLDE and obtained an admissible character or character-like solution. In some cases, the character-like solution does constitute the characters of a known CFT and hence our words are justified; in other cases, our words are not justified, especially in situations where the solution is a hitherto unknown one there is more work to be done, before we declare it  a CFT.

This paper is organised as follows. In section \ref{2s} we give technical details pertaining to MLDEs and the calculations needed to achieve classification of RCFTs. Section \ref{3ss} contains the classification of three-character RCFTs. We first study the  $\mathbf{[3, 0]}$ case in \ref{31ss}: the classification of up to central charge $96$ results in a large number of solutions, which are divided into five mutually exclusive and exhaustive categories, which we describe successively in \ref{311ss}, \ref{312ss}, \ref{313ss}, \ref{314ss} and \ref{315ss}. We then study the  $\mathbf{[3, 2]}$ case in \ref{32ss}.  In section \ref{6ss}, we collect some observations on the various new solutions of section \ref{31ss}, that seem to suggest that there are novel coset dual pairs of CFTs among them.  We conclude in section \ref{5ss}  where we consolidate our results and  give future directions. The appendix \ref{app1} tabulates all the character-like solutions of both $\mathbf{[3, 0]}$ and $\mathbf{[3, 2]}$ MLDEs.

\section{MLDE approach to RCFTs \label{2s}}

\subsection{Modular Linear Differential Equations \label{21s}}
The characters of a RCFT are a set of $n$ functions, $\chi_i(\tau)$, on the moduli space of the torus such that the torus partition function of the RCFT can be written as: 
\be \label{1n}
Z(\tau,\bar{\tau})=\sum_{i,j=0}^{n-1} M_{ij}\chi_i(\tau)\chi_j(\bar{\tau})
\ee
The characters have a $q$-expansion:
\be \label{2n}
\chi_i(\tau)=\sum_{n=0}^\infty f_n^{(i)} q^{\alpha_i+n},\qquad i = 0, 1,\cdots,n-1
\ee
where $q=e^{2\pi i\tau}$ and $f_n^{(i)}$ are the Fourier coefficients of the $i$th character. Here $\alpha_i$ are the exponents,
\be \label{10n}
\alpha_i = h_i-\frac{c}{24}
\ee
where $h_i$ are the conformal dimensions of the primaries and $c$ is the central charge. Note that, a single character can correspond to multiple primaries. The modular invariance of the  torus partition function,
\be \label{3n}
Z(\gamma \tau,\gamma\bar{\tau}) = Z(\tau,\bar{\tau}),\quad \gamma=
\begin{pmatrix}
a~ & b\\ c~ & d
\end{pmatrix}
\in \mathbf{SL(2, Z)}
\ee
implies that the characters transform as vector-valued modular forms:
\be \label{4n}
\chi_i(\gamma\tau)= \sum_kV_{ik}(\gamma)\chi_k(\tau)
\ee
where $V_{ik}$ are unitary matrices in the $n$-dimensional representation of $\mathbf{SL(2, Z)}$.

The characters of a RCFT are solutions to an ordinary differential equation. The appropriate notion of derivative is that of the  Serre-Ramanujan derivative.  We denote the upper half plane by $\mathbb{H}$,
\begin{align} \label{5n}
\mathbb{H} = \{\tau\in\mathbb{C} \ | \ \text{Im}(\tau)>0\}. 
\end{align}
The {\it Serre-Ramanujan} derivative operator (see, for example, section 2.8 of \cite{kilford} and exercise 5.1.8 of \cite{rammurthy1}) is defined as follows:
\begin{align} \label{6n}
\mathcal{D}_k\, =  12 \, q \frac{\partial}{\partial q} - k \cdot E_2
\end{align}
where $E_2(\tau)$ is the weight $2$ Eisenstein series. The operator $\mathcal{D}_k$  is a linear operator which maps weight $k$ modular objects to weight $k+2$ modular objects and satisfies the Leibniz rule. This is the differential operator with which one sets up  modular linear differential equations.

With this we can write the most general $n^{th}$ order MLDE \cite{Mathur:1988na}:
\begin{align}
\mathcal{D}^n \chi_i + \sum_{r=0}^{n-1} \phi_r(\tau)\mathcal{D}^r \chi_i = 0, \label{7n}    
\end{align}
where the solutions $\chi_i$ are the characters of a $n$-character RCFT.  In the above, the coefficient functions $\phi_r(\tau)$ are modular w.r.t the full modular group $\mathbf{SL(2,Z)}$ (even if the characters $\chi_i(\tau)$ themselves need not be, being modular w.r.t some congruence subgroup of $\mathbf{SL(2,Z)}$) and are given by rational functions of the Eisenstein series $E_4(\tau)$ and $E_6(\tau)$. Just as for differential equations with ordinary derivatives, the co-efficient functions can be expressed in terms of the Wronskians of the solutions. Define,
\begin{align}
W_r = \left(
\begin{array}{cccccc}
    \chi_1 & \chi_2  & \cdots & \cdots & \cdots &  \chi_{n} \\
    \mathcal{D}\chi_1 & \mathcal{D}\chi_2  & \cdots & \cdots & \cdots & \mathcal{D}\chi_{n} \\
    \vdots & \ddots & \ddots & \ddots & \ddots & \vdots \\
    \mathcal{D}^{r-1}\chi_1 & \mathcal{D}^{r-1}\chi_2  & \cdots & \cdots & \cdots & \mathcal{D}^{r-1}\chi_{n} \\
    \mathcal{D}^{r+1}\chi_1 & \mathcal{D}^{r+1}\chi_2  & \cdots & \cdots & \cdots & \mathcal{D}^{r+1}\chi_{n} \\
    \vdots & \ddots & \ddots & \ddots & \ddots & \vdots \\
    \mathcal{D}^n\chi_1 & \mathcal{D}^n\chi_2  & \cdots & \cdots & \cdots &  \mathcal{D}^n\chi_{n} \\
\end{array}
\right), \label{11n}
\end{align}
the Wronskian, $W$, of \eqref{11n} is given by $W\equiv W_n$. The coefficient functions of \eqref{7n} and the Wronskians \eqref{11n} are related by:
\begin{align}
\phi_r = (-1)^{n-r}\frac{W_r}{W}. \label{phir}    
\end{align}

Now let us analyse the weights of various objects in \eqref{7n}. $\chi_i$s being characters of a RCFT are weight $0$ modular functions. The first term, $\mathcal{D}^n \chi_i$, is of weight $2n$ as each operation of $\mathcal{D}$ increases the weight by $2$. Similarly  $\mathcal{D}^r \chi_i$ is of weight $2r$ and hence $\phi_r(\tau)$ are modular functions of weight $2(n-r)$. Note that since $W$ can have zeros in $\mathbb{H}$, so $\phi_r$s are not modular forms but rather are modular functions\footnote{Modular forms are holomorphic in $\mathbb{H}$ but modular functions, in general, are meromorphic in $\mathbb{H}$.}. Also, note that $W$ is a modular function of weight $n(n-1)$.

\subsubsection{The Valence Formula and the Wronskian Index}
For a modular function $f$ of weight $k$, $\nu_{\tau_0}(f)$ denotes the order of pole/zero of $f$ at $\tau=\tau_0$. Let us now look at the valence formula (see, for example, section 3.1 of \cite{kilford}) which states that,
\begin{align}
\nu_{\iota\infty}(f) + \frac{1}{2}\nu_\iota(f) + \frac{1}{3}\nu_\omega(f) + \sum^{'}_{\tau\neq \iota,\omega; \ \tau\in\mathcal{F}} \nu_\tau(f) = \frac{k}{12}, \label{VF}   
\end{align}
where $\omega$ is one of the cube root of unity, $\mathcal{F}$ is the Fundamental Domain and the $'$ above the summation means that the sum excludes $\tau\in\mathbb{H}$ which have $\text{Re}(\tau)=\frac{1}{2}$ or, which have both $|\tau|=1$ and $\text{Re}(\tau)>0$.\\
\\
Now, at $\tau\rightarrow\iota\infty$, $\chi_i\sim q^{\alpha_i}$ and hence, $W\sim q^{\sum_{i=0}^{n-1}\alpha_i}$. Hence, $W$ has a $-\sum_{i=0}^{n-1}\alpha_i$ order pole at $\iota\infty$. The last three terms on the left hand side of \eqref{VF} together can be written as  $\frac{l}{6}$ where $l$ is a non-negative integer. Hence, the valence formula now reads
\begin{align}
\sum_{i=1}^{n-1}\alpha_i + \frac{l}{6} = \frac{n(n-1)}{12}  . \label{RR}    
\end{align}
Now using \eqref{10n}, we have 
\be \label{windex}
\frac{l}{6} = \frac{n(n - 1)}{12} + \frac{n\,c}{24} - \sum_i h_i.
\ee
$l$, which is some information about the zeroes of the Wronskian (at the cusp points and the interior of the fundamental domain), is the Wronskian index of the RCFT. This Wronskian index $l$, which is expressed purely in terms of the RCFT data (as seen in \eqref{windex}), proves to be important for the classification of RCFTs.

\subsection{$\mathbf{[2,0]}$ MLDE} \label{22ss}
In this subsection, we will illustrate the theory with the example of $\mathbf{[2,0]}$ that first appeared in the classic paper \cite{Mathur:1988na}.  What we do in the following is slightly different from \cite{Mathur:1988na}, but it has the advantage that it can be generalised to the three-character story in section \ref{3ss}. Also thus we will recall the classification of $\mathbf{[2,0]}$ CFTs which show up in multiple ways in the classification of both $\mathbf{[3,0]}$ and $\mathbf{[3,2]}$ CFTs that we take up in section \ref{3ss}.

 Any $\mathbf{2}$-character MLDE has the form:
\bea \label{213}
\mathcal{D}^2 \chi_i + \phi_1(\tau)~ \mathcal{D} \chi_i + \phi_0(\tau)~ \chi_i  = 0 
\eea
with $\phi_1(\tau)$ modular of weight $2$ and $\phi_0(\tau)$ modular of weight $4$. From \eqref{phir}, 
$\phi_1(\tau) = - \frac{W_1}{W}, ~\phi_0(\tau) =  \frac{W_0}{W}$. For vanishing Wronskian index $\mathbf{l} = 0$, $W$ has no zeroes and $\phi_1(\tau), \phi_0(\tau)$ are $\mathbf{SL(2, Z)}$ modular forms (i.e. holomorphic) of weights $2$ and $4$ respectively.  Hence they are respectively $0$ and $E_4$. Thus we see how the $\mathbf{[n,l]}$ values determine the MLDE in the $\mathbf{[2,0]}$ case to be 
\bea  \label{214}
\mathcal{D}^2 \chi_i  + \mu_{1,0}~ E_4(\tau)~ \chi_i  = 0 
\eea
where $\mu_{1,0}$ is the one single parameter of this MLDE; we use a notation $\mu_{a,b}$ to be the constant parameter that multiplies the monomial $E_4(\tau)^a\,E_6(\tau)^b$. Now to obtain $\mathbf{[2,0]}$ admissible character solutions and CFTs, we plug in \eqref{2n} into \eqref{214}; at the leading order we obtain the indicial equation
\bea \label{215}
\alpha^2 - \frac{\alpha}{6} + \frac{\mu_{1,0}}{144} = 0.
\eea
We evaluate the above equation for the identity character $\alpha_0 = - \frac{c}{24}$ and use it to 
express the parameter in terms of $\alpha_0$
\bea \label{216}
\mu_{1,0} = - 144\,\alpha_0^2 + 24 \alpha_0.
\eea
Having thus converted the parameters into expressions pertaining to the identity character, we now get to the  next to leading order of \eqref{214},
\bea \label{217}
\frac{f_1}{f_0} = \frac{-576\,\alpha - 240\mu_{1,0}}{144 \alpha^2 - 264 \alpha + \mu_{1,0} + 120},
\eea
which when evaluated for the identity character and further using \eqref{216}, we obtain
\bea \label{218}
m_1 = \frac{1440 \alpha_0^2 - 264 \alpha_0}{5 + 12 \alpha_0}.
\eea
Here, $m_1$ is the Fourier coefficient (of  $q^1$) of the identity character and is a non-negative integer. Define 
\be \label{220}
N = - 120 \alpha_0
\ee
and \eqref{218} becomes 
\bea \label{221}
N^2 + N (m_1 + 22) - 50 \, m_1 = 0
\eea
a quadratic equation for $N$ with integer coefficients and by the integral root theorem we can conclude that $N$ is an integer. $-120 \alpha_0$ is an integer means that $5c$ is an integer. This is  the stage at which the problem becomes ``finite'': for the finite range $0 < c \leq 1$, there are only five possible values viz. $\frac15, \frac25, \frac35, \frac45, \frac55$.  Then we go back to \eqref{218} and rewrite in terms of $c$ to get $m_1 = \frac{5 c^2 + 22 c}{10 - c}$ and since the left hand side is positive as well as the numerator of the right hand side, the denominator needs to be positive which implies $c < 10$. Hence we find that the allowed values of the central charge are the $49$ values $\frac15, \frac25, \ldots \frac{48}{5}, \frac{49}{5}$ or $ 1 \leq N \leq 49$.

Now we compute higher Fourier coefficients. The Frobenius method gives the Fourier coefficients of a character as  rational functions of the index and the parameter, which on using \eqref{216} and the fact that $\mathbf{l} = 0$, become rational functions of the central charge. We then just compute for each of the $49$ cases and eliminate those cases for which the co-efficients are negative. It turns out that there are $10$ admissible character solutions, given in  table \ref{t0}.
\begin{table}[h] 
\begin{center}
\begin{threeparttable}
\caption{$\mathbf{[2,0]}$ }\label{t0}
\rowcolors{2}{Mygrey}{Mywhite}
\begin{tabular}{cc||c}
\hline
\hline
$N$ &  $c$ & Theory \\
\hline
2 &  $\frac25$ & $\mathcal{M}(5,2)$ \\
5 & 1 & $(\mathbf{\hat{A}_{1}})_1$  \\
10 & 2 & $(\mathbf{\hat{A}_{2}})_1$  \\
14 & $\frac{14}{5}$ & $(\mathbf{\hat{G}_{2}})_1$  \\
20 & 4 & $(\mathbf{\hat{D}_{4}})_1$  \\
26 & $\frac{26}{5}$ & $(\mathbf{\hat{F}_{4}})_1$  \\
30 & 6 & $(\mathbf{\hat{E}_{6}})_1$  \\
35 & 7 & $(\mathbf{\hat{E}_{7}})_1$  \\
38 & $\frac{38}{5}$ & $\mathbf{E_{7\frac12}}$  \\
40 &  8 & $(\mathbf{\hat{E}_{8}})_1$ \\
\hline
\hline
\end{tabular} 
\end{threeparttable}
\end{center}
\end{table}

Of these, the first eight admissible character solutions correspond to CFTs. The first entry needs an explanation. What one obtains is two indices $-\frac{1}{60}$ and $\frac{11}{60}$. The way the discussion has been set up, we equate the smaller index to be $-\frac{c}{24}$ and the bigger one to be $h-\frac{c}{24}$. This gives us a unitary CFT interpretation since we obtain positive $c$ and $h$ ($\frac25$ and $\frac15$ respectively). But in reality $\mathcal{M}(5,2)$ is a non-unitary CFT; this is also consistent with the obtained indices. If make the opposite identification, i.e.  equate the bigger index $\frac{11}{60}$ to be $-\frac{c}{24}$ and the smaller index $-\frac{1}{60}$ to be 
$h-\frac{c}{24}$ we would arrive at the correct numbers for $\mathcal{M}(5,2)$ viz. $c = - \frac{22}{5}$ and $h = -\frac15$.  From only the MLDE solution, one cannot decide if it is a unitary CFT or not. Here in table \ref{t0}, it turns out that only the first entry is a non-unitary CFT and the other CFT entries (two to eight) are all unitary CFTs.  In section \ref{3ss} we have many solutions that do not seem to be associated with any known CFTs; in all such cases, we tabulate the data as if they are unitary CFTs (just as  in table \ref{t0}), but one should keep open the possibility that they could well be non-unitary CFTs (as it happens for the first entry of table \ref{t0}).

The ninth entry corresponding to $N=38$ is not associated with any CFT  but it is a genuine admissible character and seems to be associated with some other mathematical structure (see \cite{JL:2006}). 

The last admissible character solution (for $N = 40$) is strictly not a two-character CFT. $(\mathbf{\hat{E}_{8}})_1$ is actually a one-character CFT; it is a $\mathbf{[1,2]}$ CFT.  The single character of this CFT turns out to be the identity character of the above $N = 40$ solution. The other character of the solution is an ``unstable'' character. An unstable character is a character with positive rational coefficients, but there is no positive integer available, which when multiplied makes all the Fourier coefficients integers. The positive integer needed keeps increasing as we increase the order to which we compute. This is in contrast to usual characters, which in this context may be termed ``stable'' characters, where usually after a certain order of computation one finds a positive integer that when multiplied makes all Fourier coefficients integers not only up to that order but also to all subsequent larger orders of computation. This is the first instance when a CFT with a lower number of characters shows up as a solution to a MLDE with a bigger number of characters along with unstable characters. In our study of solutions to $\mathbf{[3,0]}$ MLDEs, we will find in \ref{312ss} many two and one character CFTs that  appear as solutions accompanied with one or two unstable characters. 

The analysis that is presented here in this section (\ref{22ss}) to derive the famous MMS classification table \ref{t0} is not the same as the one that appears in the classic paper \cite{Mathur:1988na}. They arrive at a finite number of central charges ($22$) by requiring that the discriminant of \eqref{221} be a perfect square; they then compute the higher Fourier coefficients for these cases and impose positivity to obtain the classification in table \ref{t0}. The analysis presented here has the advantage that a slight generalisation  works for the three-character MLDEs that we study later in this paper.

\subsection{Inhomogeneous MLDEs} \label{23ss}
 
 The MLDEs that appear in the classification scheme being studied here \eqref{7n} and that originated with \cite{Mathur:1988na} are homogenous differential equations.  In this subsection, we develop inhomogeneous modular linear differential equations. We will show that for every $\mathbf{[n,l]}$  MLDE there is an associated inhomogeneous MLDE of order $\mathbf{n-1}$ which is equivalent to it. That is,  the solutions to the homogenous MLDE also solve  the associated inhomogeneous MLDE.  This point of view for MLDE solutions helps explain some of the solutions that we obtain in section \ref{313ss} while solving $\mathbf{[3,0]}$ homogenous MLDEs. 

The starting point is Abel's identity \cite{abel} variously referred to as Abel's formula and Abel's differential equation identity. Abel's identity allows the computation of the Wronskian of the solutions of a differential equation without having to solve the differential equation. The identity gives a first order 
 differential equation for the Wronskian; the coefficient functions for this first order differential equation are given by the coefficient functions of the differential equation under study. This theory, given by Abel  for differential equations on the complex plane, generalises to the setting of modular differential equations with Serre-Ramanujan derivatives. First, from \eqref{11n} we obtain $\mathcal{D} W = W_{n-1}$, which when combined with \eqref{phir} gives the first order differential equation for the Wronskian:
\be \label{222}
\mathcal{D} W +  \phi_{n-1}\,W = 0 .
\ee
We can solve this differential equation and obtain the Wronskians (see table \ref{tw} below for the general case and table \ref{tst} for the Wronskians of $\mathbf{[2,0]}$ RCFTs). We thus have the situation where we first have only the Wronskian index, some partial information of the Wronskian, which determines the coefficient function  $\phi_{n-1}$ of the MLDE and which in turn via \eqref{222} determines the Wronskian completely. For a $\mathbf{[n,0]}$ MLDE, $\phi_{n-1}$ vanishes and hence it's Wronskian is a constant w.r.t the Serre-Ramanujan derivative. Serre-Ramanujan constants w.r.t  the $\mathcal{D}_k$ operator can be determined by solving using a Frobenius like method:
\bea \label{223}
\mathcal{D}_k\, f = 0 \qquad \Longrightarrow \qquad f(q) = f_0~ \eta(q)^{2k}
\eea
and  are appropriate powers of the Dedekind-eta function, so as to have weight $k$. We give the Wronskians for a few MLDEs in table \ref{tw}, which includes all the particular MLDEs we study in this paper. The $A_1$ that appears in the last column is the single integration constant resulting from solving the first order differential equation \eqref{222}. $A_1$ can take different values for different CFTs; we illustrate this in table \ref{tst} where we give the $A_1$ value for the MMS $\mathbf{[2,0]}$ CFTs.
\begin{table}[h] 
\begin{center}
\begin{threeparttable}
\caption{\bf Wronskians}\label{tw}
\rowcolors{2}{Mygrey}{Mywhite}
\begin{tabular}{c||c||c}
\hline
\hline
\textbf{MLDE} &  $\phi_{n-1}$  & $W$ \\
\hline
$\mathbf{[n,0]}$ &  $0$& $A_1~\eta(q)^{2 n (n-1)}$ \\
$\mathbf{[n,2]}$ & $\frac{4 E_6}{E_4}$ & $A_1~\eta(q)^{2 n (n-1)} \left[ \frac{E_4}{\eta(q)^{8}} \right]$  \\
$\mathbf{[n,3]}$ &  $\frac{6 E_4^2}{E_6}$ & $A_1~\eta(q)^{2 n (n-1)} \left[ \frac{E_6}{\eta(q)^{12}} \right]$ \\
$\mathbf{[n,4]}$ &  $\frac{8 E_6}{E_4}$ & $A_1~\eta(q)^{2 n (n-1)}  \left[ \frac{E_4}{\eta(q)^{8}} \right]^2$ \\
\hline
\hline
\end{tabular} 
\end{threeparttable}
\end{center}
\end{table}

\begin{table}[h] 
\begin{center}
\begin{threeparttable}
\caption{$\mathbf{[2,0]}$ \textbf{Wronskians} }\label{tst}
\rowcolors{2}{Mygrey}{Mywhite}
\begin{tabular}{c||c||c}
\hline
\hline
$c$ & Theory & Wronskian \\
\hline
$\frac25$ & $\mathcal{M}(5,2)$ & $-\frac{1}{5}\,\eta(q)^{4}$ \\
1 & $(\mathbf{\hat{A}_{1}})_1$ & $\frac{1}{2}\,\eta(q)^{4}$ \\
2 & $(\mathbf{\hat{A}_{2}})_1$ & $\eta(q)^{4}$ \\
$\frac{14}{5}$ & $(\mathbf{\hat{G}_{2}})_1$ & $\frac{14}{5}\,\eta(q)^{4}$ \\
4 & $(\mathbf{\hat{D}_{4}})_1$ & $4\,\eta(q)^{4}$ \\
$\frac{26}{5}$ & $(\mathbf{\hat{F}_{4}})_1$ & $\frac{78}{5}\,\eta(q)^{4}$ \\
6 & $(\mathbf{\hat{E}_{6}})_1$ & $18\,\eta(q)^{4}$ \\
7 & $(\mathbf{\hat{E}_{7}})_1$ & $42\,\eta(q)^{4}$ \\
\hline
\hline
\end{tabular} 
\end{threeparttable}
\end{center}
\end{table}

Solving \eqref{222}  is effectively one integration. And we are left with $\mathbf{n-1}$ integrations more to be performed. These are done by solving the following $\mathbf{n-1}$-order MLDE, which is obtained as follows.  Consider the formula for the Wronskian; we will illustrate this for $\mathbf{[3,0]}$:
\bea \label{224}
A_1 \eta^{12} =  \mathcal{D}^2\chi_3 \left[ \chi_1\,\mathcal{D}\chi_2 - \chi_2\,\mathcal{D}\chi_1 \right] + \mathcal{D}\chi_3 \left[ \chi_2\,\mathcal{D}^2\chi_1 - \chi_1\,\mathcal{D}^2\chi_2 \right] + \chi_3 \left[ \mathcal{D}\chi_1\mathcal{D}^2\chi_2 - \mathcal{D}\chi_2\,\mathcal{D}^2\chi_1  \right]  
\nonumber \\
\eea
and rewrite it as
\bea
\mathcal{D}^2\chi_1 + \mathcal{D}\chi_1\left[ \frac{\chi_3\mathcal{D}^2\chi_2 - \chi_2\mathcal{D}^2\chi_3}{\chi_2\mathcal{D}\chi_3 - \chi_3\mathcal{D}\chi_2} \right] + \chi_1 \left[ \frac{\mathcal{D}\chi_2\mathcal{D}^2\chi_3 - \mathcal{D}\chi_3\mathcal{D}^2\chi_2}{\chi_2\mathcal{D}\chi_3 - \chi_3\mathcal{D}\chi_2} \right]  = \frac{A_1 \eta^{12}}{\chi_2\mathcal{D}\chi_3 - \chi_3\mathcal{D}\chi_2} \label{225}
\eea
In general, we obtain an inhomogeneous MLDE of order $\mathbf{n-1}$, the inhomogeneous term given by the Wronskian from \eqref{222}. The Wronskian is a known function at this stage and appears in \eqref{225}, if in addition, we consider $\chi_1$ and  $\chi_2$ as given, we can read \eqref{225} as a differential equation for $\chi_3$. We then solve this first by obtaining a particular solution. This can be done again with a Frobenius-like procedure. Let us denote the particular solution by  $A_1 P(q)$. The general solution to \eqref{225} is then given by a linear combination of the particular solution and the homogenous solutions. The homogenous solutions of \eqref{225} are nothing but $\chi_2$ and $\chi_3$. Hence the general solution is 
\bea \label{226}
\chi_1(q) = A_1\, P(q) + A_2 \, \chi_2(q) + A_3\, \chi_3(q).
\eea
Let us denote the Fourier expansion of the particular solution and the Fourier expansions of the other characters by
\bea 
A_1 P(q) = q^{\frac{1}{2} - \alpha_2 - \alpha_3}\sum\limits_{n=0}^{\infty} g^{(1)}_n q^n, \quad \chi_2(q) = q^{\alpha_2}\sum\limits_{n=0}^{\infty} g^{(2)}_n q^n, \quad \chi_3(q) = q^{\alpha_3}\sum\limits_{n=0}^{\infty} g^{(3)}_n q^n. \label{227}
\eea
Then we have
\bea \label{228}
\chi_1(q) = q^{\frac{1}{2} - \alpha_2 - \alpha_3}\sum\limits_{n=0}^{\infty} \left[g^{(1)}_n + A_2\, q^{-\frac{1}{2} + 2 \alpha_2 + \alpha_3}\,  g^{(2)}_n  + A_3\, q^{-\frac{1}{2} +  \alpha_2 + 2 \alpha_3}\,  g^{(3)}_n   \right] q^n.
\eea
If the values of $\alpha_2$ and $\alpha_3$ for which character-like solutions exist are such that $-\frac{1}{2} + 2 \alpha_2 + \alpha_3$ and $-\frac{1}{2} +  \alpha_2 + 2 \alpha_3$  are not non-negative integers, then to get character-like solution for $\chi_1$, we have to set $A_2$ and $A_3$ to be zero. This is what happens in most examples: the first character is nothing but the particular solution of the inhomogeneous MLDE. But one can imagine the following situation. 

(i) If  $-\frac{1}{2} + 2 \alpha_2 + \alpha_3$ is a non-negative integer, then $A_2$ isn't required to vanish. $A_2$ can take any positive integral value and we would have a good character solution for $\chi_1$.
\begin{align} \label{229}
&-\frac{1}{2} + 2 \alpha_2 + \alpha_3 \in \mathbf{Z_{\geq 0}}, \qquad  A_2 \in \mathbf{Z_{\geq 0}}\nonumber \\ 
&\chi_1(q) = q^{\frac{1}{2} - \alpha_2 - \alpha_3}\sum\limits_{n=0}^{\infty} \left[g^{(1)}_n + A_2\, q^{-\frac{1}{2} + 2 \alpha_2 + \alpha_3}\,  g^{(2)}_n    \right] q^n.
\end{align}
We thus have an infinite number of admissible  character solutions, parametrised by $A_2$, in  \eqref{229}. All members of this infinite family have the same indices and hence the same $c$, $h_1$, $h_2$ and also they have the same Wronskian. But they are different solutions (they differ in  the first character).

(ii) If  $-\frac{1}{2} + \alpha_2 + 2 \alpha_3$ is a non-negative integer, then $A_3$ isn't required to vanish. $A_3$ can take any positive integral value and we would have a good character solution for $\chi_1$.
\begin{align} \label{230}
&-\frac{1}{2} +  \alpha_2 + 2 \alpha_3 \in \mathbf{Z_{\geq 0}}, \qquad  A_3 \in \mathbf{Z_{\geq 0}}\nonumber \\ 
&\chi_1(q) = q^{\frac{1}{2} - \alpha_2 - \alpha_3}\sum\limits_{n=0}^{\infty} \left[g^{(1)}_n + A_3\, q^{-\frac{1}{2} +  \alpha_2 + 2 \alpha_3}\,  g^{(3)}_n    \right] q^n.
\end{align}
We thus have an infinite number of admissible  character solutions, parametrised by $A_3$, in  \eqref{230}. All members of this infinite family have the same indices and hence the same $c$, $h_1$, $h_2$ and also they have the same Wronskian. But they are different solutions (they differ in  the first character).

In our study of character-like solutions to $\mathbf{[3,0]}$ MLDEs we will encounter two infinite families of CFTs (in section \ref{313ss}) each family having the same $c$, $h_1$, $h_2$ values, one following \eqref{229} and another \eqref{230}.

\section{A Classification of three-character CFTs }\label{3ss}

In this section, we study three character CFTs. First we study three character CFTs with a vanishing Wronskian index i.e. $\mathbf{[3,0]}$ CFTs and subsequently we study three character CFTs with a Wronskian index equalling two i.e. $\mathbf{[3,2]}$ CFTs. We have been made aware by the recent concurrent work \cite{Kaidi:2021ent} that for three character MLDEs new admissible character solutions are expected when the Wronskian index is a multiple of $3$.

\subsection{Classifying $\mathbf{[3,0]}$ CFTs}\label{31ss}

Any $\mathbf{3}$-character MLDE has the form:
\bea \label{31q}
\mathcal{D}^3 \chi_i + \phi_2(\tau)~\mathcal{D}^2 \chi_i + \phi_1(\tau)~ \mathcal{D} \chi_i + \phi_0(\tau)~ \chi_i  = 0 
\eea
with $\phi_2(\tau)$, $\phi_1(\tau)$ and $\phi_0(\tau)$ modular of weight $2$, $4$ and $6$ respectively.  From \eqref{phir}, 
$\phi_1(\tau) = - \frac{W_1}{W}, ~\phi_0(\tau) =  \frac{W_0}{W}$. For vanishing Wronskian index $\mathbf{l} = 0$, $W$ has no zeroes and $\phi_2(\tau)$, $\phi_1(\tau)$, and $\phi_0(\tau)$ are $\mathbf{SL(2, Z)}$ modular forms (i.e. holomorphic) of weights $2$, $4$ and $6$ respectively.  Hence they are respectively $0$, $E_4$ and $E_6$. Thus we see how the $\mathbf{[n,l]}$ values determine the MLDE in the $\mathbf{[3,0]}$ case to be 
\bea \label{54s}
\mathcal{D}^3 \chi_i  + \mu_{1,0}\,E_4 \, \mathcal{D}\chi_i  + \mu_{0,1}\,E_6 \, \chi_i= 0, 
\eea
This is a two parameter MLDE.  Let us denote the three characters by $\chi_0(q), \chi_1(q), \chi_2(q)$ where $\chi_0(q)$ is the identity character. The characters have the $q$-expansion $ \chi_i(q)= q^{\alpha_i}\sum\limits_{n=0}^{\infty} f^{(i)}_n q^n$  $~f^{(i)}_0 \neq 0$ and each $f^{(i)}_n$ a positive integer.  $\alpha_i$ is referred to as the index of the character. The three indices are named such that $\alpha_0 < \alpha_1 < \alpha_2$.  The three rational numbers that define a three-character RCFT viz. the central charge $c$,  the conformal dimensions of the non-identity characters $h_1$ and $h_2$  with $h_1 < h_2$, are taken to be related to the indices : $\alpha_0 = - \frac{c}{24}, \alpha_1 = h_1 - \frac{c}{24}, \alpha_2 = h_2 - \frac{c}{24}$.\footnote{In the following, we will be searching for solutions with a negative $\alpha_0$. This means that the assignment $\alpha_0 = - \frac{c}{24}, \alpha_1 = h_1 - \frac{c}{24}, \alpha_2 = h_2 - \frac{c}{24}$ together with $\alpha_0 < \alpha_1 < \alpha_2$ implies that we are searching for solutions that give unitary CFTs. It may seem that we will miss out on non-unitary CFTs. But it turns out non-unitary CFTs also show up as solutions within this scheme. We can illustrate this with the example of the non-unitary Virasoro minimal model $\mathcal{M}(7,2)$ for which we have $c = - \frac{68}{7}, h_1 = - \frac27, h_2 = -\frac37$ which corresponds to indices $\alpha_0 = \frac{17}{42}, \alpha_1 = \frac{5}{42}, \alpha_2 = -\frac{1}{42}$ which are not ordered according to $\alpha_0 < \alpha_1 < \alpha_2$. But if we rename the indices $\overline{\alpha_0} = -\frac{1}{42}, \overline{\alpha_1} = \frac{5}{42}, \overline{\alpha_2} = \frac{17}{42}$ so that they are correctly ordered $\overline{\alpha_0} < \overline{\alpha_1} < \overline{\alpha_2}$ and define central charges and conformal dimensions $\overline{\alpha_0} = - \frac{\overline{c}}{24}, \overline{\alpha_1} = \overline{h_1} - \frac{\overline{c}}{24}, \overline{\alpha_2} = \overline{h_2} - \frac{\overline{c}}{24}$ we get $\overline{c} = \frac47, \overline{h_1} = \frac17, \overline{h_2} = \frac27$ which looks like a unitary CFT. And this is how the non-unitary CFT $\mathcal{M}(7,2)$ shows up within our scheme of computation (see the second row of table \ref{t7}). It turns out \cite{Castro-Alvaredo:2017udm} that the central charge $\overline{c}$ has a physical meaning as the central charge that controls  renormalisation group flows between non-unitary CFTs (it is called $\mathbf{c_{eff}}$ in \cite{Castro-Alvaredo:2017udm}). Another non-unitary CFT that shows up in our tables is $\mathcal{M}(5,2)^{\otimes 2}$ which has $\mathbf{c_{eff}} = \frac45$ (see third row of table \ref{t7}). 

Thus among the many solutions that we obtain in this paper which are completely unknown or not yet fully characterised as CFTs, there could be non-unitary CFTs also. And we would have  tabulated their $\mathbf{c_{eff}}$.\label{f3}}

A linear differential equation, when solved by a Frobenius-like method, determines the ratios $\frac{f^{(i)}_n}{f^{(i)}_0}$. For the identity character,  $f^{(0)}_0 = 1$; in the following we will write $f^{(0)}_n \equiv m_n, n \geq 1$ and all the $m_n$'s are non-negative integers. For the other  characters, $i = 1, 2$,  $f^{(i)}_0$  need not be $1$ and and hence in our search for character-like solutions we will require that the ratios determined by the differential equation are all positive rational numbers such that upon multiplying by a suitable positive integer all of them become positive integers. This step depends on the  order $n$ up to which one solves the differential equation. We will denote this positive integer by $D_1$ and $D_2$ for the characters $\chi_1$ and $\chi_2$ respectively. $f^{(1)}_0$ and $f^{(2)}_0$ cannot be determined by the method pursued here which is the Frobenius-like method for solving linear differential equations; they can be any multiple of $D_1$ and $D_2$ respectively.

At the leading order in the $q$-expansion, the differential equation \eqref{54s} gives the following indicial equation
\bea \label{55s}
\alpha^{3} - \frac{\alpha^{2}}{2} + \left(\frac{\mu_{1,0}}{144} + \frac{1}{18}\right)\alpha + \frac{\mu_{0,1}}{1728} = 0,
\eea
whose solutions are the indices of the characters: $\alpha_0, \alpha_1, \alpha_2$. At the next to leading order in the $q$-expansion, the differential equation \eqref{54s} gives the following equation
\bea\label{56s}
\frac{f_1}{f_0} = \frac{ 42 \mu_{0,1} - 240 \mu_{1,0} \alpha - 192 \alpha - 1728 \alpha^2 }{ 80 +  \mu_{1,0} + 288 \alpha + 432 \alpha^2  }.
\eea
We now evaluate the two equations \eqref{54s} and \eqref{55s} for the identity character and interpret these equations as two equations for the two parameters which give the solution
\bea \label{57s}
\mu_{0,1} &=& -1728 \alpha_0^{3} + 864\alpha_0^{2} - 12  \mu_{1,0} \alpha_0 - 96 \alpha_0 \nonumber \\
\mu_{1,0} &=& -\frac{72576\,\alpha_0^{3} + (432\,m_1 - 34560)\alpha_0^{2} + (288\,m_1 + 4224)\alpha_0 + 80\,m_1}{m_1 + 744\,\alpha_0}.
\eea
We have thus determined the two parameters of the MLDE in terms of objects associated with the identity character viz. $\alpha_0$ and $m_1$. We now consider the differential equation \eqref{54s} at the next order in the $q$-expansion, which when evaluated for the identity character, is an equation for $m_2$. After using \eqref{57s} we obtain a polynomial equation in the variables $\alpha_0$, $m_1$ and $m_2$. We observe that if we define a new variable 
\be \label{58s} 
N \equiv -1680\,\alpha_0
\ee
the polynomial equation is the following one with indeterminates $N$, $m_1$ and $m_2$.
\begin{align}
&N^4 + (m_1^2 + 93\,m_1 - 2\,m_2 + 955)N^3 + (-2380\,m_1^{2} + 28770\,m_1 + 7700\,m_2 + 167160)N^2 \nonumber \\
&+ (1372000\,m_1^2 + 10760400\,m_1 - 7330400\,m_2 - 9800\,m_1 m_2)N \nonumber \\ 
&+ 13720000\,m_1 m_2 - 329280000\,m_1^2 = 0.  \label{59s}
\end{align}

If we consider \eqref{59s} as a polynomial equation to determine $N$, which at this stage is a rational number, we find that the coefficients are all integers. Given that the coefficient of $N^4$ is $1$, which is the reason why we choose the definition \eqref{58s} in the first place, we can use the integer-root-theorem to conclude that $N$ has to be an integer. This is a important turning point in our analysis. $N$ which was just rational is now restricted to be an integer. This means that the central charge, which is given by  $c = \frac{N}{70} $ is a rational number whose denominator (not necessarily in the $\frac{p}{q}$ form with $p,q$ coprime) is always $70$. This also means that the central charge when expressed in it's irreducible form $\frac{p}{q}$ with $p,q$ coprime, $q$ can take values $1, 2, 5, 7, 10, 14, 35$ and $70$. More significantly, our problem has become a ``finite'' problem in the sense that if one is searching for CFTs in a certain range of central charge, say $0 < c \leq 1$, one only needs to do $70$ computations for $c = \frac{1}{70}, \frac{2}{70}, \frac{3}{70}, \ldots \frac{69}{70}, \frac{70}{70}.$  Furthermore, \eqref{59s} is a equation with all indeterminates being positive integers, hence a Diophantine equation. We pick a value of $N$, or equivalently $c$, and look for positive integer solutions to \eqref{59s}. The considerations that will give us the rest of the procedure for solving \eqref{54s} are as follows. 

The two parameters of the MLDE given in \eqref{57s}, can now (\eqref{58s}) can be thought of as functions of $N$ and $m_1$. The indicial equation \eqref{55s}  becomes a cubic equation whose coefficients are functions of $N$ and $m_1$ only. Furthermore we readily know one of the roots viz. $\alpha_0$ or more pertinently  $- \frac{N}{1680} $. Thus the cubic indicial equation factorises into linear and quadratic equations:
\begin{align} \label{60s}
&(1680 \alpha + N)\left[(8749440 N - 19756800 m_1)\alpha^2 + (-5208 N^2 + 11760\,m_1 N - 4374720  N + 9878400\,m_1)\alpha \right. \nonumber \\ 
& \left. + \, N^{3} + 14\,m_1 N^{2} + 924\,N^{2} - 29400\,m_1N + 141120\, N + 9878400\,m_1\right] = 0. 
\end{align}
The roots of the quadratic equation determine the other two indices $\alpha_1$ and $\alpha_2$ and eventually $h_1$ and $h_2$. For rational roots the discriminant has to be the square of a rational number, but since the discriminant is an integer (on account of all the coefficients being integers), it needs to be a perfect square to be able to result in rational roots. If we denote the discriminant by $k^2$ with $k$ chosen to be positive, we have 
\begin{align} 
&3457440000\,m_1^{2} - 1657376000\,m_1 N - 8232000\,m_1^{2} N + 55899200 N^{2}  \nonumber \\
&+ 3528000\,m_1 N^{2} + 4900\,m_1^{2} N^{2} 
+ 52080\,N^{3} - 2100\,m_1 N^{3} - 31 N^{4} = k^{2}. \label{61s}
\end{align}
We note that \eqref{61s} is an equation where all the indeterminates viz. $N$, $m_1$ and $k$ are positive integers and hence a Diophantine equation.  The roots are then given by 
\begin{eqnarray} \label{62s}
\frac{58800\,m_1 - 26040\,N + 70\,m_1\,N - 31\,N^{2} \pm 3\,k}{3360(70\,m_1 - 31\,N)}
\end{eqnarray}
with $\alpha_1$ the smaller root and the $\alpha_2$ the bigger one.

The procedure\footnote{\label{f4}Most of the key equations \eqref{59s}, \eqref{60s}, \eqref{61s} for the solution of the $\mathbf{[3,0]}$ MLDE can already be found in an appendix of the early paper on this subject \cite{Mathur:1988gt}. We  have been able to use that theory to come up with this procedure that involves a finite number of computations for a finite range of central charge.} to obtain character like solutions to \eqref{54s} can now be stated. 

(i) We solve the two Diophantine equations \eqref{59s} and \eqref{61s} simultaneously. We fix a value of $N$ and then obtain solutions to \eqref{59s}. Each such solution is a set of positive integers for $N$, $m_1$ and $m_2$.  We take the $m_1$ and $N$ values of each such solution and see if we can find positive integer values for $k$ that satisfy \eqref{61s}. At the end of this procedure, we will have a set of four positive integers, for $N$, $m_1$, $m_2$ and $k$ which simultaneously solve \eqref{59s} and \eqref{61s}. For example, for $0 < N \leq 70$, we obtain six simultaneous solutions to \eqref{59s} and \eqref{61s} given in Table \ref{t1}.
\begin{table}[h] 
\begin{center}
\begin{threeparttable}
\caption{Simultaneous solutions to Diophantine equations for $0 < N \leq 70$}\label{t1}
\begin{tabular*}{\textwidth}{@{\extracolsep{\fill}}cccc}
\hline
\hline
\makebox[0pt][l]{\fboxsep0pt\colorbox{Mywhite} {\strut\hspace*{\linewidth}}}
$N$  & $m_1$ & $m_2$ & $k$    \\
\hline
\makebox[0pt][l]{\fboxsep0pt\colorbox{Mywhite} {\strut\hspace*{\linewidth}}}
28  & 1 & 1 & 67032 \\
\makebox[0pt][l]{\fboxsep0pt\colorbox{Mygrey} {\strut\hspace*{\linewidth}}}
35  & 0 & 1 & 265825 \\
\makebox[0pt][l]{\fboxsep0pt\colorbox{Mywhite} {\strut\hspace*{\linewidth}}}
40  & 1 & 2 & 187200  \\
\makebox[0pt][l]{\fboxsep0pt\colorbox{Mygrey} {\strut\hspace*{\linewidth}}}
56  & 2 & 3 & 178752   \\
\makebox[0pt][l]{\fboxsep0pt\colorbox{Mywhite} {\strut\hspace*{\linewidth}}}
70  & 1 & 4 & 441000    \\
\makebox[0pt][l]{\fboxsep0pt\colorbox{Mygrey} {\strut\hspace*{\linewidth}}}
70  & 3 & 4 & 137200    \\
\hline
\hline
\end{tabular*} 
\end{threeparttable}
\end{center}
\end{table}

(ii) At this stage, we have the three indices: $\alpha_0$, $\alpha_1$ and $\alpha_2$ from equations \eqref{58s} and \eqref{62s} which further means that we have the $c$, $h_1$ and $h_2$ of the putative CFTs. But we have imposed positivity on the Fourier coefficients of the character only upto order $q^2$ and only for the identity character. Hence, now in this next stage of the procedure, we impose positivity. That is we determine the Fourier coefficients of all the characters upto some high order (for us $q^{2000}$), which are afterall functions of the index of the character and the parameters in the MLDE which in turn are functions of the known $\alpha_0$ and $m_1$ \eqref{57s} and check if they are positive. We reject solutions to the Diophantine equations which do not survive the positivity constraints. We keep all the other solutions and in the process also compute $D_1$ and $D_2$, the positive integers that when multiplied to $\chi_1$ and $\chi_2$, makes all their Fourier coefficients to be positive integers as well. For example, the $N=28$ solution in the first row of Table \ref{t1} turns out to have negative Fourier coefficients for the identity character and hence is rejected.

We implement the just delineated procedure for $ N \leq 6720 $ or in other words for $c \leq 96$. Our results are tabulated in  appendix \ref{app1}. Each row gives details of a character-like solution to the $\mathbf{[3,0]}$ MLDE \eqref{54s}. At this initial stage of analysis, we think of them as character-like solutions and not yet CFTs. Every row consists, in the first four columns, of the (simultaneous) solution to the Diophantine equations \eqref{59s}, \eqref{61s}. In the next two columns, we give the $D_1$ and $D_2$ values (not for solutions that will eventually turn out to be well-known CFTs, for example the ones that occur in the partial classification of \cite{Das:2020wsi}). The next three columns, that is the seventh, eighth and the ninth columns, consists of the $c$, $h_1$ and $h_2$ values and the reader who is only interested in the results of this paper can focus on just these columns (with the reminder of the unitary CFT interpretation see footnote \ref{f3}).  The last column consists of an assignment of the solution to one of five mutually exclusive and exhaustive categories: $\mathbf{I},$ $\mathbf{II},$ $\mathbf{III},$ $\mathbf{IV}$ and $\mathbf{V}$, which we describe below. Our computations for $ N \leq 6720 $ is given in the eight tables \ref{t7} - \ref{t14} of appendix \ref{app1}. These eight tables together comprise $227$ rows of character-like solutions. Solutions with central charges between two consecutive integers are grouped together and are given between two horizontal double lines.  If we were to tabulate all our computations, we would have had $307$ rows. We find a pattern for the last $80$ rows (and beyond) and hence don't put it here. Each row, given in appendix \ref{app1}, represents one character-like solution, except four rows, which each represent multiple solutions and together comprise two infinite classes of solutions. Of the $227$ rows of solutions, the split into the five categories $\mathbf{I},$ $\mathbf{II},$ $\mathbf{III},$ $\mathbf{IV}$ and $\mathbf{V}$ are $122, 22, 61, 15,$ and $7$ respectively.  If we had given all the $307$ rows, the split into the five categories $\mathbf{I},$ $\mathbf{II},$ $\mathbf{III},$ $\mathbf{IV}$ and $\mathbf{V}$ would have been  $202, 22, 61, 15,$ and $7$ respectively.

\subsubsection{Category $\mathbf{I}$ } \label{311ss}

The first category of solutions are the known RCFTs with three characters and vanishing Wronskian index. Recently, a computation of the $\mathbf{[n,l]}$ values for a large class of known RCFTs was made and tabulated in \cite{Das:2020wsi}. The results of that study can be viewed as a partial classification of RCFTs for a given $\mathbf{[n,l]}$; the complete classification would come from a study of all solutions to the associated  $\mathbf{[n,l]}$ MLDE.  We will recall the partial classification for $\mathbf{[3,0]}$ CFTs from that paper (after correcting a minor typo) here:
\bea\label{63s}
&&(\mathbf{\hat{A}_1})_2,\qquad (\mathbf{\hat{A}_3})_1,\qquad (\mathbf{\hat{A}_4})_1, \nonumber \\
&& (\mathbf{\hat{B}_r})_1 \quad \text{for all} \quad r \geq 3, \nonumber \\
&&(\mathbf{\hat{C}_2})_1, \nonumber \\
&& (\mathbf{\hat{D}_r})_1 \quad \text{for all} \quad r \geq 5, \nonumber \\
&&(\mathbf{\hat{E}_8})_2, \nonumber \\
&&\mathcal{M}(7,2), \qquad \mathcal{M}(4,3), \nonumber \\
&&\mathcal{M}(5,2)\otimes\mathcal{M}(5,2), \qquad (\mathbf{\hat{A}_1})_1 \otimes (\mathbf{\hat{A}_1})_1,\qquad (\mathbf{\hat{A}_2})_1 \otimes (\mathbf{\hat{A}_2})_1, \qquad (\mathbf{\hat{D}_4})_1 \otimes (\mathbf{\hat{D}_4})_1, \nonumber \\ 
&&(\mathbf{\hat{E}_6})_1 \otimes (\mathbf{\hat{E}_6})_1,\qquad (\mathbf{\hat{E}_7})_1 \otimes (\mathbf{\hat{E}_7})_1, \qquad (\mathbf{\hat{F}_4})_1 \otimes (\mathbf{\hat{F}_4})_1,\qquad (\mathbf{\hat{G}_2})_1 \otimes (\mathbf{\hat{G}_2})_1.
\eea
In the tables  \ref{t7} - \ref{t14} of appendix \ref{app1}, we have $122$ solutions in category $\mathbf{I}$. We will collect them into table \ref{t2} and table \ref{t3} here. The $N$ and $m_1$ values of any row uniquely identifies that row. The reader can correlate a row of table \ref{t2} or table \ref{t3} with a (unique) row from  tables \ref{t7} - \ref{t14} of appendix \ref{app1} by comparing the $N$ and $m_1$ values. 
\begin{table}[h] 
\begin{center}
\begin{threeparttable}
\caption{\bf Category I : irreducible }\label{t2}
\rowcolors{2}{Mygrey}{Mywhite}
\begin{tabular}{cc||c}
\hline
\hline
$N$ &  $m_1$ & Theory \\
\hline
35 &  0 & $\mathcal{M}(4,3)$ \\
40 & 1 & $\mathcal{M}(7,2)$  \\
70 &  1 & $(\mathbf{\hat{so}(2)})_1$ \\
105 &  3 & $(\mathbf{\hat{A}_{1}})_2$ \\
175 &  10 & $(\mathbf{\hat{C}_{2}})_1$ \\
210 &  15 & $(\mathbf{\hat{A}_{3}})_1$  \\
245 &  21 & $(\mathbf{\hat{B}_{3}})_1$ \\
280 & 24 & $(\mathbf{\hat{A}_{4}})_1$ \\
1085 &  248 & $(\mathbf{\hat{E}_{8}})_2$ \\
70r+35 & $\frac{2r(2r+1)}{2}$ & $(\mathbf{\hat{B}_{r}})_1$ \\
70(r+1) & $\frac{2(r+1)(2r+1)}{2}$ & $(\mathbf{\hat{D}_{r+1}})_1$ \\
\hline
\hline
\end{tabular} 
\begin{tablenotes}
\small
\item where $4\leq r\leq 55$
\end{tablenotes}
\end{threeparttable}
\end{center}
\end{table}

Table \ref{t2} contains  $113$ of the $122$ solutions of category $\mathbf{I}$. Each of them corresponds to an irreducible three-character CFT, that is, it is not a tensor-product of theories with fewer characters. We can locate them in the partial classification \eqref{63s} in the first five rows. The third row of table \ref{t2} is not there in \eqref{63s}. This is a known CFT, corresponding to two decoupled copies of the Ising model. It is also a WZW CFT, $(\mathbf{\hat{so}(2)})_1$, and can perhaps be included in the $(\mathbf{\hat{D}_r})_1$ series. 

We performed computations up to $N \leq 6720$ or for $c \leq 96$. But we are reporting only computations up to $N \leq 3920$ or for $c \leq 56$, see table \ref{t13}. In the range $56 < c \leq 96$, we have $80$ admissible character solutions and all of them are in category  $\mathbf{I}$. They are  $(\mathbf{\hat{B}_{r}})_1$ for $56 \leq r \leq 95$ snd $(\mathbf{\hat{D}_{r}})_1$ for $57 \leq r \leq 96$.  It is tempting to conjecture that for all $c  > 96$ also there are only category $\mathbf{I}$ solutions (the $\mathbf{B}$ and $\mathbf{D}$ series WZW CFTs ) and attempt to find a proof for the same.

\begin{table}[h] 
\begin{center}
\begin{threeparttable}
\caption{\bf Category I : tensor}\label{t3}
\rowcolors{2}{Mygrey}{Mywhite}
\begin{tabular}{c||cc||c}
\hline
\hline
S. No. & $N$ &  $m_1$ & Theory \\
\hline
1. & 56 &  2 & $\mathcal{M}(5,2)^{\otimes 2}$ \\
2. & 140 & 6 & $(\mathbf{\hat{A}_{1}})_1^{\otimes 2}$  \\
3. & 280 &  16 & $(\mathbf{\hat{A}_{2}})_1^{\otimes 2}$  \\
4. & 392 &  28 & $(\mathbf{\hat{G}_{2}})_1^{\otimes 2}$ \\
5. & 560 &  56 & $(\mathbf{\hat{D}_{4}})_1^{\otimes 2}$ \\
6. & 728 &  104 & $(\mathbf{\hat{F}_{4}})_1^{\otimes 2}$  \\
7. & 840 &  156 & $(\mathbf{\hat{E}_{6}})_1^{\otimes 2}$ \\
8. & 980 & 266 & $(\mathbf{\hat{E}_{7}})_1^{\otimes 2}$ \\
9. & 1064 &  380 & $\mathbf{E_{7\frac{1}{2}}}^{\otimes 2}$ \\
\hline
\hline
\end{tabular} 
\end{threeparttable}
\end{center}
\end{table}

Table \ref{t3} contains the other $9$ of the $122$ solutions of category $\mathbf{I}$. Each of them corresponds to the tensor product of two copies of a $\mathbf{[2,0]}$ character-like solution (see table \ref{t0}). Eight of these can be located in the partial classification of CFTs \eqref{63s} in the last two rows. The ninth $\mathbf{E_{7\frac{1}{2}}}^{\otimes 2}$ is not a CFT, but very much an admissible character solution \cite{JL:2006}.

Of all the solutions in category $\mathbf{I}$, two of them are non-unitary CFTs, the second and first entries of tables \ref{t2} and \ref{t3} respectively. See footnote \ref{f3}.

\subsubsection{Category $\mathbf{II}$ }\label{312ss}

Even at the outset, it should already be stated that all the solutions in category $\mathbf{II}$, except possibly two, are not to be considered as admissible-character solutions that will give consistent CFTs. This is because while all the three characters satisfy the positivity constraints, there are one or more characters that are ``unstable''.  An unstable character is a character with positive rational coefficients, but there is no positive integer available, which when multiplied makes all the Fourier coefficients integers. The positive integer needed keeps increasing as we increase the order to which we compute. This is in contrast to usual characters, which in this context may be termed ``stable'' characters, where usually after a certain order of computation one finds a positive integer that when multiplied makes all Fourier oefficients integers not only up to that order but also to all subsequent larger orders of computation.

In the original MMS classification for $\mathbf{[2,0]}$ CFTs \cite{Mathur:1988na}, given here in table \ref{t0}, of the ten solutions that pass the positivity criteria, only nine of them are admissible character solutions with only ``stable'' characters. Eight of them (the first eight in table \ref{t0}) are genuine CFTs and one of them (the ninth entry of table \ref{t0}), is related to another mathematical structure \cite{JL:2006}. But that paper does include a solution (the last entry of table \ref{t0}), one of whose characters is an ``unstable'' character and the other ``stable'' character is the character of the $(\mathbf{\hat{E}_{8}})_1$ CFT which is a $\mathbf{[1,2]}$ CFT.  Our category $\mathbf{II}$ solutions are being reported here, keeping in line with this aspect of original MMS classification table. 

It is an interesting question to ask: which two-character CFTs along with one ``unstable'' character are solutions to $\mathbf{[3,0]}$ MLDEs? There is also the related question: which one-character CFTs along with two ``unstable'' characters are solutions to $\mathbf{[3,0]}$ MLDEs? The table \ref{t4} answers these questions.

The solutions of category $\mathbf{II}$ are related to $\mathbf{[n,l]}$ CFTs or $\mathbf{[n,l]}$ admissible character solutions with $\mathbf{n} \leq 2$ and $\mathbf{l} \leq 4$. Two solutions viz. S. No. 9 and 10 of table \ref{t4}, have two ``unstable'' characters and the stable character is that of a $\mathbf{[1,2]}$ and $\mathbf{[1,4]}$ CFT, respectively. $18$ solutions in table \ref{t4} have one ``unstable'' character with the ``stable'' characters that of known two-character CFTs. The $\mathbf{[2,0]}$ CFT $\mathcal{M}(5,2)$ is not there in table \ref{t4}. It turns out that there is a solution to the $\mathbf{[3,0]}$ MLDE  \eqref{54s}, two of whose linearly independent solutions are the characters of $\mathcal{M}(5,2)$ and the other is a solution with negative Fourier coefficients at low orders (reminiscent of quasi-characters \cite{Mukhi:2020gnj}). This occurs for the  $N = 28, m_1 = 1, m_2 = 1, k =  67032$ solution to the Diophantine equations. But we do not report this here in the tables. 

The last two solutions left to discuss are curious ones: S. No.4  and S. No.15 of table \ref{t4}. In each case all characters are ``stable'' characters. For S. No.4, two of the characters are the characters of the $(\mathbf{\hat{D}_4})_1$ CFT and the third character is the non-identity character of the $(\mathbf{\hat{D}_4})_1$ CFT. Thus this solution has the curious feature that two of the three characters are identical. This means  that Wronskian vanishes, which does not contradict the fact that all $\mathbf{[3,0]}$ solutions should have $A_1 \eta^{12}$ as the Wronskian (see section \ref{23ss}), if  $A_1 = 0$ (which is also a possibility). It is not clear to us if we should consider this as a new $\mathbf{[3,0]}$ admissible character or just derived from an old $\mathbf{[2,0]}$ solution. The solution in S. No.15 is similar, except that the two-character CFT is a $\mathbf{[2,2]}$ CFT that  appeared in \cite{Hampapura:2015cea}.

\begin{table}[h] 
\begin{center}
\begin{threeparttable}
\caption{\bf Category II}\label{t4}
\rowcolors{2}{Mygrey}{Mywhite}
\begin{tabular}{l||cc||l}
\hline
\hline
S.No. & $N$ & $m_1$ & \textbf{Identification}  \\
\hline
1. & 70 & 3 & {\bf (2,0)}: \ $(\mathbf{\hat{A}_1})_1$ \\
2. & 140 &  8 & {\bf (2,0)}: \ $(\mathbf{\hat{A}_2})_1$ \\
3. & 196 &  14 & {\bf (2,0)}: \ $(\mathbf{\hat{G}_2})_1$ \\
4. & 280 &  28 & {\bf (2,0)}: \ $(\mathbf{\hat{D}_4})_1$ \\
5. & 364 &  52 & {\bf (2,0)}: \ $(\mathbf{\hat{F}_4})_1$ \\
6. & 420 &  78 & {\bf (2,0)}: \ $(\mathbf{\hat{E}_6})_1$ \\
7. & 490 &  133 & {\bf (2,0)}: \ $(\mathbf{\hat{E}_7})_1$ \\
8. & 532 &  190 & {\bf (2,0)}: \ $\mathbf{E_{7\frac{1}{2}}}$ \\
9. & 560 &  248 & {\bf (1,2)}: \ $(\mathbf{\hat{E}_{8}})_1$ \\
10. & 1120 & 496 & {\bf (1,4)}: \ $(\mathbf{\hat{E}_{8}})_1^{\otimes 2}$ \\
11. & 1148 &  410 & {\bf (2,2)}: \ CFT of \cite{Hampapura:2015cea} \\
12 & 1190 &  323 & {\bf (2,2)}: \ GHM-dual of $(\mathbf{\hat{E}_{7}})_1$ \\
13. & 1260 &  234 & {\bf (2,2)}: \ GHM-dual of $(\mathbf{\hat{E}_{6}})_1$ \\
14. & 1316 &  188 & {\bf (2,2)}: \ GHM-dual of $(\mathbf{\hat{F}_{4}})_1$ \\
15. & 1400 &  140 & {\bf (2,2)}: \ GHM-dual of $(\mathbf{\hat{D}_{4}})_1$ \\
16. & 1484 &  106 & {\bf (2,2)}: \ GHM-dual of $(\mathbf{\hat{G}_{2}})_1$ \\
17. & 1540 &  88 & {\bf (2,2)}: \ GHM-dual of $(\mathbf{\hat{A}_{2}})_1$ \\
18. & 1610 &  69 & {\bf (2,2)}: \ GHM-dual of $(\mathbf{\hat{A}_{1}})_1$ \\
19. & 1652 &  59 & {\bf (2,2)}: \ CFT of \cite{Hampapura:2015cea} \\
20. & 2268 &  4 & {\bf (2,4)} CFT of \cite{Chandra:2018pjq} \\
21. & 2310 &  3 & {\bf (2,4)} CFT of \cite{Chandra:2018pjq} \\
22. & 2380 &  1 & {\bf (2,4)} CFT of \cite{Chandra:2018pjq} \\
\hline
\hline
\end{tabular} 
\end{threeparttable}
\end{center}
\end{table}

\subsubsection{Category $\mathbf{III}$ }\label{313ss}

In their paper \cite{FM:2019}, C. Franc and G. Mason studied certain kinds of vertex operator algebras of rank $3$, which seem to be related to $\mathbf{[3,0]}$ CFTs. The similarity is that they also study (certain kinds of)
character-like solutions to the $\mathbf{[3,0]}$ MLDE \eqref{54s} that we are studying in this paper. We have rediscovered all these Franc-Mason solutions here in our work. Some of the Franc-Mason solutions pertain to well-known (even prior to \cite{FM:2019}) CFTs. There are Franc-Mason solutions that occur within our category $\mathbf{I}$, category $\mathbf{II}$ as well as category $\mathbf{IV}$ (see below). The rest of the Franc-Mason solutions find their place in the present category $\mathbf{III}$, There are $61$ rows in the eight tables \ref{t7}-\ref{t14} of appendix \ref{app1}. $57$ of these rows correspond to $57$ CFTs; the rest  four rows each correspond to multiple CFTs which we discuss below.

Now we discuss the first of the left over four rows of category $\mathbf{III}$. This is in table \ref{t8} and occurs at $N = 560, m_1 = d_1, m_2 = d_2,  k = k_1$, where $d_1$, $d_2$, $k_1$ are integers: $0\leq d_1\leq 247$, $d_2=156+16d_1$ and $k_1=4860800-19600d_1$. Note that these are $248$ different CFTs all with the same central charge and even same scaling dimensions. The second of the left over rows of  category $\mathbf{III}$ is in table \ref{t8} and occurs at $N = 560, m_1 = e_1, m_2 = e_2,  k = k_2$ where $e_1$, $e_2$, $k_2$ are integers: $e_1\geq 249$, $e_2=156+16e_1$ and $k_2=-4860800+19600e_1$. Note that these are an infinite class of CFTs all with the same central charge and even same scaling dimensions. These two classes of CFTs comprise the first infinite class of Franc-Mason solutions \cite{FM:2019} ($(x=0,y=-1/2)$). We can understand these infinite solutions all with the same central charge and scaling dimensions from an aspect of MLDE solutions that we derived in section \ref{23ss}, equation \eqref{230}: translating to the notation of that equation, we have $\alpha_1 = -\frac13, \alpha_2 = \frac16,  \alpha_3 = \frac23 \Longrightarrow -\frac{1}{2} +  \alpha_2 + 2 \alpha_3 = 1$.

Now we discuss the third of the left over four rows of category $\mathbf{III}$. This is in table \ref{t9} and occurs at $N = 1120, m_1 = d_1, m_2 = d_2,  k = k_1$, where $d_1$, $d_2$, $k_1$ are integers:$0\leq d_1\leq 495$, $d_2=2296+136d_1$ and $k_1=9721600-19600d_1$. Note that these are $496$ different CFTs all with the same central charge and even same scaling dimensions. The fourth of the left over rows of  category $\mathbf{III}$ is in table \ref{t9} and occurs at $N = 1120, m_1 = e_1, m_2 = e_2,  k = k_2$ where $e_1$, $e_2$, $k_2$ are integers: $e_1\geq 497$, $e_2=2296+136e_1$ and $k_2=-9721600+19600e_1$. Note that these are an infinite class of CFTs all with the same central charge and even same scaling dimensions. These two classes of CFTs comprise the second infinite class of Franc-Mason solutions \cite{FM:2019}($(x=1,y=0)$). We can understand these infinite solutions all with the same central charge and scaling dimensions from an aspect of MLDE solutions that we derived in section \ref{23ss}, equation \eqref{229}: translating to the notation of that equation, we have $\alpha_1 = -\frac23, \alpha_2 = \frac13,  \alpha_3 = \frac56 \Longrightarrow -\frac{1}{2} +  2\alpha_2 +  \alpha_3 = 1$.

\subsubsection{Category $\mathbf{IV}$ }\label{314ss}

The paper \cite{Gaberdiel:2016zke} gives a new kind of coset construction for CFTs, which is sometimes referred to as a ``novel coset construction'' (to perhaps distinguish it from WZW cosets). In this construction one starts with a meromorphic CFT ($\mathcal{H}$) and quotients it by an affine sub-theory ($\mathcal{D}$), also referred to as the denominator theory and produces the coset CFT ($\mathcal{C}$). Schematically one writes $\mathcal{C} = \mathcal{H} / \mathcal{D}$ and refers to the $\mathcal{C}$-theory as the novel coset dual of the $\mathcal{D}$-theory. We will refer to these dual theories as ``GHM-duals'' in this paper. 

We have $15$ solutions in category $\mathbf{IV}$ in the tables \ref{t7}-\ref{t14} of appendix \ref{app1}. Each of these are GHM-dual CFTs. All of them belong to only one particular case of GHM-duality among the many that are studied in  \cite{Gaberdiel:2016zke}. This is the case where both the denominator CFT $\mathcal{D}$ as well as the GHM-dual CFT $\mathcal{C}$ are $\mathbf{[3,0]}$ CFTs. It is convenient and suggestive to use the notation $\tilde{\mathcal{D}}$ for $\mathcal{C}$. The central charges and the scaling dimensions of the two CFTs, in this particular case of GHM-duality, are related as follows:
\bea \label{11s}
c^{\mathcal{D}} + c^{\tilde{\mathcal{D}}} = 24, \qquad h_1^{\mathcal{D}} + h_1^{\tilde{\mathcal{D}}} = 2, \qquad h_2^{\mathcal{D}} + h_2^{\tilde{\mathcal{D}}} = 2. 
\eea

We collect all the solutions in category $\mathbf{IV}$ here in table \ref{t5}. Each of the GHM-dual CFTs are duals of  category $\mathbf{I}$ CFTs. From our computations, and from \eqref{11s} which comprise necessary conditions,  we can only identify potential GHM-dual CFTs.  Whether it is actually a GHM-dual CFT requires more analysis involving meromorphic CFTs \cite{Schellekens:1993}, \cite{Chandra:2018ezv}; which is beyond the scope of the present work. But all the CFTs in our table \ref{t5} are actual GHM-dual CFTs and the meromorphic CFT associated to each and other details can be found in table 2 of \cite{Gaberdiel:2016zke}.

\begin{table}[h] 
\begin{center}
\begin{threeparttable}
\caption{\bf Category IV}\label{t5}
\rowcolors{2}{Mygrey}{Mywhite}
\begin{tabular}{l||cc||l}
\hline
\hline
S.No. & $N$ &  $m_1$ & \textbf{Identification}  \\
\hline
1. & 1050 &  255 & GHM-dual of $(\mathbf{\hat{D}_{9}})_1$ \\
2. & 1190 &  221 & GHM-dual of $(\mathbf{\hat{D}_{7}})_1$ \\
3. & 1225 & 210 & GHM-dual of $(\mathbf{\hat{B}_{6}})_1$ \\
4. & 1260 &  198 & GHM-dual of $(\mathbf{\hat{D}_{6}})_1$ \\
5. & 1295 &  185 & GHM-dual of $(\mathbf{\hat{B}_{5}})_1$ \\
6. & 1330 &  171 & GHM-dual of $(\mathbf{\hat{D}_{5}})_1$ \\
7. & 1365 &  156 & GHM-dual of $(\mathbf{\hat{B}_{4}})_1$ \\
8. & 1400 &  120 & GHM-dual of $(\mathbf{\hat{A}_{4}})_1$ \\
9. & 1435 &  123 & GHM-dual of $(\mathbf{\hat{B}_{3}})_1$ \\
10. & 1470 &  105 & GHM-dual of $(\mathbf{\hat{A}_{3}})_1$ \\
11. & 1505 & 86 & GHM-dual of $(\mathbf{\hat{C}_{2}})_1$ \\
12. & 1575 &  45 & GHM-dual of $(\mathbf{\hat{A}_{1}})_2$ \\
13 & 1645 &  0 & GHM-dual of $\mathcal{M}(4,3)$ \\
14. & 2296 &  0 & GHM-dual of $\mathcal{M}(5,2)^{\otimes 2}$ \\
15. & 2360 &  0 & GHM-dual of $\mathcal{M}(7,2)$ \\
\hline
\hline
\end{tabular} 
\end{threeparttable}
\end{center}
\end{table}

\subsubsection{Category $\mathbf{V}$ }\label{315ss}

In the last category, from the tables \ref{t7}-\ref{t14} we have seven rows corresponding to seven admissible character solutions, which we have gathered in table \ref{t6}.  These are new solutions. They do not seem to fall in any of the known classes of solutions. At this stage of the analysis, one can only claim to have obtained new character-like solutions, also referred to as admissible character solutions. They are not yet CFTs; to establish that we need more work \cite{wp}. We can only say that we potentially 
have \emph{seven new $\mathbf{[3,0]}$ CFTs}.

\begin{table}[h] 
\begin{center}
\begin{threeparttable}
\caption{\bf Category V}\label{t6}
\rowcolors{2}{Mygrey}{Mywhite}
\begin{tabular}{l||cc||ccc}
\hline
\hline
S.No. & $N$ &  $m_1$ & $c$ & $h_1$ & $h_2$  \\
\hline
1. & 840 &  318 & 12 & 1/3 & 5/3 \\
2. & 1400 &  80 & 20 & 4/3 & 5/3 \\
3. & 1400 &  728 & 20 & 1/3 & 8/3 \\
4. & 1400 &  890 & 20 & 2/3 & 7/3 \\
5. & 1960 &  1948 & 28 & 2/3 & 10/3 \\
6. & 2520 &  3384 & 36 & 2/3 & 13/3 \\
7. & 3080 &  3146 & 44 & 1/3 & 17/3 \\
\hline
\hline
\end{tabular} 
\end{threeparttable}
\end{center}
\end{table}

Here, we present the Fourier expansion (a few terms) of the seven potentially new $\mathbf{[3,0]}$ RCFTs.
\begin{itemize}
\item[]{{\bf 1. c = 12} 
\begin{enumerate}
\item $\chi_0(q) = q^{-\frac{1}{2}}\left(1 + 318q + 8514q^2 + 126862q^3 + 1269771q^4 + 9695754q^5 + \mathcal{O}(q^6)\right)$ 
\item $\chi_{\frac{1}{3}}(q) = q^{-\frac{1}{6}}\left(1 + 178q + 4634q^2 + 61924q^3 + 566277q^4 + 4041864q^5 + \mathcal{O}(q^6)\right)$ 
\item $\chi_{\frac{5}{3}}(q) = q^{\frac{7}{6}}\left(1 + 23q + 272q^2 + 2286q^3 + 15318q^4 + 87091q^5 + \mathcal{O}(q^6)\right)$ 
\end{enumerate}} 

\item[]{{\bf  2. c = 20} 
\begin{enumerate}
\item $\chi_0(q) = q^{-\frac{5}{6}}\left(1 + 80q + 46790q^2 + 2654800q^3 + 68308625q^4 + 1122213936q^5 + \mathcal{O}(q^6)\right)$ 
\item $\chi_{\frac{4}{3}}(q) = q^{\frac{1}{2}}\left(5 + 840q + 34398q^2 + 744040q^3 + 10930320q^4 + 122840280q^5 + \mathcal{O}(q^6)\right)$
\item $\chi_{\frac{5}{3}}(q) = q^{\frac{5}{6}}\left(4 + 355q + 11240q^2 + 209810q^3 + 2791180q^4 + 29171789q^5 + \mathcal{O}(q^6)\right)$ 
\end{enumerate}} 

\item[]{{\bf 3.  c = 20} 
\begin{enumerate}
\item $\chi_0(q) = q^{-\frac{5}{6}}\left(1 + 728q + 106406q^2 + 3894424q^3 + 82707185q^4 + 1242579288q^5 + \mathcal{O}(q^6)\right)$ 
\item $\chi_{\frac{1}{3}}(q) = q^{-\frac{1}{2}}\left(2 + 717q + 83124q^2 + 3031214q^3 + 62776974q^4 + 904657032q^5 + \mathcal{O}(q^6)\right)$
\item $\chi_{\frac{8}{3}}(q) = q^{\frac{11}{6}}\left(1 + 40q + 806q^2 + 11076q^3 + 117599q^4 + 1031680q^5 + \mathcal{O}(q^6)\right)$ 
\end{enumerate}} 

\item[]{{\bf 4.  c = 20} 
\begin{enumerate}
\item $\chi_0(q) = q^{-\frac{5}{6}}\left(1 + 890q + 55700q^2 + 2695300q^3 + 68460905q^4 + 1122686166q^5 + \mathcal{O}(q^6)\right)$ 
\item $\chi_{\frac{2}{3}}(q) = q^{-\frac{1}{6}}\left(5 + 728q + 58000q^2 + 1822700q^3 + 33995325q^4 + 452188800q^5 + \mathcal{O}(q^6)\right)$
\item $\chi_{\frac{7}{3}}(q) = q^{\frac{3}{2}}\left(10 + 423q + 9180q^2 + 134925q^3 + 1516500q^4 + 13956435q^5 + \mathcal{O}(q^6)\right)$ 
\end{enumerate}} 

\item[]{{\bf 5.  c = 28} 
\begin{enumerate}
\item $\chi_0(q) = q^{-\frac{7}{6}}\left(1 + 1948q + 424724q^2 + 23967552q^3 + 1047902951q^4 + 31801518220q^5 + \mathcal{O}(q^6)\right)$ 
\item $\chi_{\frac{2}{3}}(q) = q^{-\frac{1}{2}}\left(25 + 9084q + 862632q^2 + 59416816q^3 + 2267443743q^4 + 55824667884q^5 + \mathcal{O}(q^6)\right)$
\item $\chi_{\frac{10}{3}}(q) = q^{\frac{13}{6}}\left(11 + 628q + 18439q^2 + 368372q^3 + 5618761q^4 + 69789060q^5 + \mathcal{O}(q^6)\right)$ 
\end{enumerate}}

\item[]{{\bf 6. c = 36} 
\begin{enumerate}
\item $\chi_0(q) = q^{-\frac{3}{2}}\left(1 + 3384q + 1337850q^2 + 186743064q^3 + 10611868761q^4 + 442989392472q^5 + \mathcal{O}(q^6)\right)$ 
\item $\chi_{\frac{2}{3}}(q) = q^{-\frac{5}{6}}\left(2 + 1294q + 243835q^2 + 19695524q^3 + 1210616824q^4 + 50488594386q^5 + \mathcal{O}(q^6)\right)$
\item $\chi_{\frac{13}{3}}(q) = q^{\frac{17}{6}}\left(4 + 289q + 10688q^2 + 268714q^3 + 5154084q^4 + 80349657q^5 + \mathcal{O}(q^6)\right)$ 
\end{enumerate}}

\item[]{{\bf 7. c = 44} 
\begin{enumerate}
\item $\chi_0(q) = q^{-\frac{11}{6}}\left(1 + 3146q + 1906130q^2 + 467324061q^3 + 61502739069q^4 + 5175295282540q^5 + \mathcal{O}(q^6)\right)$ 
\item $\chi_{\frac{1}{3}}(q) = q^{-\frac{3}{2}}\left(13 + 20178q + 8638218q^2 + 1682129028q^3 + 187246128501q^4 + \mathcal{O}(q^5)\right)$
\item $\chi_{\frac{17}{3}}(q) = q^{\frac{23}{6}}\left(19 + 1702q + 76646q^2 + 2317292q^3 + 52998283q^4 + 979386560^5 + \mathcal{O}(q^6)\right)$ 
\end{enumerate}}
\end{itemize}

We leave the further study of the above new admissible character solutions in category $\mathbf{V}$ as well as the ones in category $\mathbf{III}$ for the near future \cite{wp}, especially the question as to whether they constitute consistent CFTs. We have been informed that some work in that direction has been done in \cite{Kaidi:2021ent}.

\subsection{Classifying $\mathbf{[3,2]}$ CFTs}\label{32ss}
We now study three character rational conformal field theories with a Wronskian index given by $2$.  The characters of  every $\mathbf{[3,2]}$ CFT is a solution to the $\mathbf{[3,2]}$ MLDE :
\bea \label{12t}
\mathcal{D}^3 \chi_i  + \mu_{-1, 1}\, \frac{E_6}{E_4}\, \mathcal{D}^{2} \chi_i + \mu_{1,0}\,E_4 \, \mathcal{D}\chi_i  + \mu_{0,1}\,E_6 \, \chi_i= 0, 
\eea
At this stage there are three parameters in the MLDE, but one of them gets fixed: $\mu_{-1,1} = 4$ and we are in the same situation as the $\mathbf{[3,0]}$ case i.e. two free parameters to scan. The analysis and the strategy follows the $\mathbf{[3,0]}$ case and we will be brief here, highlighting only the contrasts. At the leading order in the $q$-expansion, the differential equation \eqref{12t} gives the following indicial equation:
\begin{eqnarray}
1728\, \alpha^{3} - 288\, \alpha^{2} + 12\, \mu_{1, 0}\, \alpha + \mu_{0, 1} = 0 \label{13t}
\end{eqnarray}
Note that the sum of the indices is $\frac16$ which is what is dictated by the Wronskian index being two from \eqref{windex} and which is a direct consequence of choosing $\mu_{-1,1} = 4$. 
At the next to leading order in the $q$-expansion, the differential equation \eqref{12t} gives the following equation
\bea \label{14t}
\frac{f_1}{f_0} = -\frac{2\left(17280 \alpha^3 - 19872 \alpha^2 + 48\alpha(5 \mu_{1,0} + 66)-11 \mu_{0, 1}\right)}{432\alpha^2 + 384\alpha + \mu_{1, 0} + 120}
\eea
We again now evaluate the two equations \eqref{13t} and \eqref{14t} for the identity character and interpret these equations as two equations for the two parameters which give the solution
\begin{eqnarray} \label{15t}
\mu_{0, 1} &=& \frac{288 \alpha_0 (\alpha_0 + 1)\left(-1440 \alpha_0^2 + 264 \alpha_0 + 12 \alpha_0 m_1 + 5 m_1\right)}{744\alpha_0 + m_1} \nonumber \\
\mu_{1, 0} &=& -\frac{24\left(3024\,\alpha_0^{3} + 6(3m_1 - 320)\alpha_0^2 + 8(2m_1 + 33)\alpha_0 + 5m_1\right)}{744\alpha_0 + m_1}.
\end{eqnarray}
We again have thus determined the two parameters of the MLDE in terms of objects associated with the identity character viz. $\alpha_0$ and $m_1$. We now consider the differential equation \eqref{12t} at the next order in the $q$-expansion, which when evaluated for the identity character, is an equation for $m_2$. After using \eqref{15t} we obtain a polynomial equation in the variables $\alpha_0$, $m_1$ and $m_2$. We observe that if we define a new variable 
\begin{eqnarray} \label{16t}
N = -5040\, \alpha_0
\end{eqnarray} 
the polynomial equation is the following one with indeterminates $N$, $m_1$ and $m_2$. 
\begin{align} \label{17t}
&N^{4} + \left(3m_1^{2} + 279\,m_1 - 6\,m_2 - 75695\right)N^{3} -\left(26460\,m_1^{2} + 2397150\,m_1 - 79380\,m_2 + 71137080\right)N^{2} \nonumber \\
&+\left(74440800\,m_{1}^{2} - 264600\, m_1 m_2 + 7888784400\,m_1 - 236728800\, m_2\right)N  \nonumber \\
 &- \left(87127488000\, m_1^{2} - 1259496000\, m_1 m_2\right) = 0  
\end{align}
We can use the integer-root-theorem to conclude that $N$ has to be an integer. This is again an important turning point in our analysis. $N$ which was just rational is now restricted to be an integer. This means that the central charge, which is given by  $c = \frac{N}{210} $ is a rational number whose denominator (not necessarily in the $\frac{p}{q}$ form with $p,q$ coprime) is always $210$. Our problem has again become a ``finite'' problem in the sense that if one is searching for CFTs in a certain range of central charge, say $0 < c \leq 1$, one only needs to do $210$ computations for $c = \frac{1}{210}, \frac{2}{210}, \frac{3}{210}, \ldots \frac{209}{210}, \frac{210}{210}.$  Equation \eqref{17t} is one of the Diophantine equations. Again we use \eqref{15t} and \eqref{16t} in the indicial equation, which then factorises into linear and quadratic equations:
\begin{eqnarray}
&&\left(\alpha + \frac{N}{5040} \right)\left((78744960\, N - 533433600\, m_1)\alpha^2 + (105840\, N m_1 + 88905600\, m_1 - 15624\, N^{2} - 13124160\, N)\alpha\right.\nonumber\\
&&\left. + 42\, N^{2}m_1 - 299880\, N m_1 + 444528000\, m_1 + N^{3} - 4116\,N^{2} - 4656960\, N\right) = 0 \label{18t}
\end{eqnarray}
The discriminant of the quadratic equation gives the second Diophantine equation:
\begin{eqnarray}
&&-31\, N^{4} - 140(45m_1 - 5332)\,N^{3} + 4900(146320 + 7152\,m_1 + 9\,m_1^{2})\,N^{2} 
-8232000\,m_1(33\,m_1 + 8092)\,N \nonumber\\
&&+ 418350240000\,m_1^2 = k^2. \label{19t}
\end{eqnarray}
The other two indices are given by 
\bea \label{20t}
 \frac{26040\, N + 31\, N^{2} - 176400\,m_1 - 210\,N m_1 \pm 3\, k}{10080(31\,N - 210\,m_1)}.
\eea
We follow the same procedure as for the $\mathbf{[3,0]}$ case. First we find positive integer solutions to the two Diophantine equations \eqref{17t} and \eqref{19t}. We do this for $N \leq 20160$ to investigate central charge in the range $0 < c \leq 96$. For every solution, we then compute further Fourier coefficients via the Frobenius-like method and reject all solutions which don't satisfy positivity criteria necessary for character-like solutions. 

Our results are tabulated in table \ref{t15} of appendix \ref{app1}. We find that there are exactly nine character-like solutions. Each of these solutions has the following feature. The character $\chi_1(q)$ is a constant character. How this comes about is that $h_1 = \frac{c}{24}$ so that $\alpha_1 = 0$ and furthermore all Fourier coefficients of this character vanish except the first one i.e $f_0^{(1)} \neq 0, f_n^{(1)} = 0, n \neq 0$. Thus $\chi_1(q) = f_0^{(1)}$ is a constant and we call this a constant character. Every solution in table \ref{t15} is thus comprised of the characters of a $\mathbf{[2,0]}$ CFT or more generally of an admissible character solution of a $\mathbf{[2,0]}$ MLDE adjoined with a constant character, 

We can show using elementary MLDE theory that this is an example of a general phenomenon. One can always adjoin a constant character to a $\mathbf{[n,l}]$ CFT to obtain a $\mathbf{[n+1,n+l}]$ CFT. What we are seeing in our results is for $\mathbf{n} = 2$ and $\mathbf{l} = 0$.

Our computations seem to suggest the following classification for $\mathbf{[3,2]}$ CFTs: \emph{The only admissible character solutions are the ones obtained by adjoining a constant character to a $\mathbf{[2,0]}$ admissible character solution.} Recent concurrent work \cite{Kaidi:2021ent} suggests that for three character MLDEs new admissible character solutions are expected only when the Wronskian index is a multiple of $3$. This is in agreement with our classification above, which does not contain any new admissible character solutions. 

\section{Potential GHM-dual pairs of $\mathbf{[3,0]}$ CFTs\label{6ss}}

Among the five categories of $\mathbf{[3,0]}$ solutions given in section \ref{31ss}, we will consider categories $\mathbf{I}$, $\mathbf{II}$ and $\mathbf{IV}$ as well-understood CFTs. Solutions in category $\mathbf{V}$ are  new and need to be studied more, as well as solutions in category $\mathbf{III}$, the Franc-Mason vertex operator algebras \cite{FM:2019}. In this section, we will tabulate some observations about solutions in these two categories and hope to present a more complete study in the future \cite{wp}.

In the novel coset construction of Gaberdiel-Hampapura-Mukhi \cite{Gaberdiel:2016zke}, the central charges and the scaling dimensions of the denominator $\mathbf{[3,0]}$ CFT and the coset $\mathbf{[3,0]}$ CFT are related by \eqref{11s}. We will define a pair of $\mathbf{[3,0]}$ CFTs $(T,\tilde{T})$  to be potentially GHM-dual, or that the theories $(T,\tilde{T})$ constitute a potential GHM-dual pair,  if their central charges and scaling dimensions follow:
\begin{align}
c+\tilde{c} = 24, \qquad h_1 + \tilde{h}_1 = 2, \qquad h_2 + \tilde{h}_2 = 2.
\end{align}
We have been able to identify $27$ potential GHM-dual pairs of CFTs, which we have tabulated in table \ref{tghm}. There are two kinds of these pairs of solutions. The first kind ($17$ pairs) is one in which one of the CFTs of the pair is a known WZW CFT and the other is a unknown CFT from category $\mathbf{III}$ or category $\mathbf{V}$. This kind of a situation is closer to the GHM construction where one of the CFTs (the denominator CFT) is a WZW CFT.  The second kind ($10$ pairs) is one where both the CFTs of the pair is an unknown CFT from the category $\mathbf{III}$. This seems very interesting. We hope to investigate and report on these potential GHM-dual pairs of CFTs in future work \cite{wp}.

\begin{table}[!hbtp] 
\begin{center}
\begin{threeparttable}
\caption{Potential GHM-dual Pairs of $\mathbf{[3,0]}$ CFTs}\label{tghm}
\rowcolors{2}{Mygrey}{Mywhite} 
\begin{tabular}{ccccc||ccccc||l}
\hline
\hline
$N$ & $m_1$ & $c$ & $h_1$ & $h_2$ & $\Tilde{N}$ & $\Tilde{m}_1$ & $\Tilde{c}$ & $\Tilde{h}_1$ & $\Tilde{h}_2$ & {\bf GHM-dual pair}  \\
\hline
70 & 1 & 1 & 1/2 & 1/8 & 1610 & 23 & 23 & 3/2 & 15/8 & $(\mathbf{\hat{so}(2)})_1$ $\leftrightarrow$ $\mathbf{III}$  \\
120 & 6 & 12/7 & 2/7 & 3/7 & 1560 & 78 & 156/7 & 12/7 & 11/7 & $\mathbf{III}$ $\leftrightarrow$ $\mathbf{III}$ \\
140 & 6 & 2 & 1/2 & 1/4 & 1540 & 66 & 22 & 3/2 & 7/4 & $(\mathbf{\hat{A}_{1}})_1^{\otimes 2}$ $\leftrightarrow$ $\mathbf{III}$ \\
168 & 3 & 12/5 & 1/5 & 3/5 & 1512 & 27 & 108/5 & 9/5 & 7/5 & $\mathbf{III}$ $\leftrightarrow$ $\mathbf{III}$ \\
280 & 16 & 4 & 2/3 & 1/3 & 1400 & 80 & 20 & 4/3 & 5/3 & $(\mathbf{\hat{A}_{2}})_1^{\otimes 2}$ $\leftrightarrow$ $\mathbf{V}$ \\
392 & 28 & 28/5 & 4/5 & 2/5 & 1288 & 92 & 92/5 & 6/5 & 8/5 & $(\mathbf{\hat{G}_{2}})_1^{\otimes 2}$ $\leftrightarrow$ $\mathbf{III}$ \\
440 & 88 & 44/7 & 4/7 & 5/7 & 1240 & 248 & 124/7 & 10/7 & 9/7 & $\mathbf{III}$ $\leftrightarrow$ $\mathbf{III}$ \\
504 & 144 & 36/5 & 3/5 & 4/5 & 1176 & 336 & 84/5 & 7/5 & 6/5 & $\mathbf{III}$ $\leftrightarrow$ $\mathbf{III}$ \\
520 & 156 & 52/7 & 4/7 & 6/7 & 1160 & 348 & 116/7 & 10/7 & 8/7 & $\mathbf{III}$ $\leftrightarrow$ $\mathbf{III}$ \\
525 & 105 & 15/2 & 1/2 & 15/16 & 1155 & 231 & 33/2 & 3/2 & 17/16 & $(\mathbf{\hat{B}_{7}})_1$ $\leftrightarrow$ $\mathbf{III}$ \\
560 & $d_1$ & 8 & 1/2 & 1 & 1120 & $\Tilde{d_1}$ & 16 & 3/2 & 1 & $\mathbf{III}$ $\leftrightarrow$ $\mathbf{III}$ \\
595 & 255 & 17/2 & 1/16 & 3/2 & 1085 & 465 & 31/2 & 31/16 & 1/2 & $\mathbf{III}$ $\leftrightarrow$ $(\mathbf{\hat{B}_{15}})_1$ \\
600 & 210 & 60/7 & 3/7 & 8/7 & 1080 & 378 & 108/7 & 11/7 & 6/7 & $\mathbf{III}$ $\leftrightarrow$ $\mathbf{III}$ \\
616 & 220 & 44/5 & 2/5 & 6/5 & 1064 & 380 & 76/5 & 8/5 & 4/5 & $\mathbf{III}$ $\leftrightarrow$ $\mathbf{III}$ \\
616 & 253 & 44/5 & 1/5 & 7/5 & 1064 & 437 & 76/5 & 9/5 & 3/5 & $\mathbf{III}$ $\leftrightarrow$ $\mathbf{III}$ \\
630 & 261 & 9 & 1/8 & 3/2 & 1050 & 435 & 15 & 15/8 & 1/2 & $\mathbf{III}$ $\leftrightarrow$ $(\mathbf{\hat{D}_{15}})_1$ \\
665 & 171 & 19/2 & 1/2 & 19/16 & 1015 & 261 & 29/2 & 3/2 & 13/16 & $(\mathbf{\hat{B}_{9}})_1$ $\leftrightarrow$ $\mathbf{III}$ \\
665 & 266 & 19/2 & 3/16 & 3/2 & 1015 & 406 & 29/2 & 29/16 & 1/2 & $\mathbf{III}$ $\leftrightarrow$ $(\mathbf{\hat{B}_{14}})_1$ \\
680 & 221 & 68/7 & 3/7 & 9/7 & 1000 & 325 & 100/7 & 11/7 & 5/7 & $\mathbf{III}$ $\leftrightarrow$ $\mathbf{III}$ \\
700 & 270 & 10 & 1/4 & 3/2 & 980 & 378 & 14 & 7/4 & 1/2 & $\mathbf{III}$ $\leftrightarrow$ $(\mathbf{\hat{D}_{14}})_1$ \\
728 & 104 & 52/5 & 3/5 & 6/5 & 952 & 136 & 68/5 & 7/5 & 4/5 & $(\mathbf{\hat{F}_{4}})_1^{\otimes 2}$ $\leftrightarrow$ $\mathbf{III}$ \\
735 & 210 & 21/2 & 1/2 & 21/16 & 945 & 270 & 27/2 & 3/2 & 11/16 & $(\mathbf{\hat{B}_{10}})_1$ $\leftrightarrow$ $\mathbf{III}$ \\
735 & 273 & 21/2 & 5/16 & 3/2 & 945 & 351 & 27/2 & 27/16 & 1/2 & $\mathbf{III}$ $\leftrightarrow$ $(\mathbf{\hat{B}_{13}})_1$ \\
770 & 231 & 11 & 1/2 & 11/8 & 910 & 273 & 13 & 3/2 & 5/8 & $(\mathbf{\hat{D}_{11}})_1$ $\leftrightarrow$ $\mathbf{III}$ \\
770 & 275 & 11 & 3/8 & 3/2 & 910 & 325 & 13 & 13/8 & 1/2 & $\mathbf{III}$ $\leftrightarrow$ $(\mathbf{\hat{D}_{13}})_1$ \\
805 & 253 & 23/2 & 1/2 & 23/16 & 875 & 275 & 25/2 & 3/2 & 9/16 & $(\mathbf{\hat{B}_{11}})_1$ $\leftrightarrow$ $\mathbf{III}$ \\
805 & 276 & 23/2 & 7/16 & 3/2 & 875 & 300 & 25/2 & 25/16 & 1/2 & $\mathbf{III}$ $\leftrightarrow$ $(\mathbf{\hat{B}_{12}})_1$ \\
\hline
\hline
\end{tabular} 
\end{threeparttable}
\end{center}
\end{table}

\newpage

\section{Conclusion and Future Directions \label{5ss}}
 
In this paper, we have directly solved two two-parameter MLDEs, the  $\mathbf{[3,0]}$ and $\mathbf{[3,2]}$ MLDEs and classified admissible character solutions and potential $\mathbf{[3,0]}$ and $\mathbf{[3,2]}$ RCFTs for the range of central charge within $0 < c \leq 96$. We have divided the $\mathbf{[3,0]}$ solutions into five mutually exclusive and exhaustive categories namely $\mathbf{I, \ II, \ III, \ IV, \ V}$. Out of these five categories, category $\mathbf{V}$ includes \emph{seven} admissible character  solutions and \emph{seven potentially new RCFTs}. We will investigate, in future, which of these seven admissible character solutions in category $\mathbf{V}$ are actually RCFTs (by checking if their fusion rule coefficients come out to be positive using the Verlinde formula); some work in this direction has appeared in \cite{Kaidi:2021ent}.  At this point, we conjecture that for $c > 96$ we would only get the WZW CFTs belonging to the $\mathbf{\hat{B}}_r$ and $\mathbf{\hat{D}}_r$ series. We have also classified $\mathbf{[3,0]}$ admissible character solutions and have discovered that each of them is a $\mathbf{[2,0]}$ admissible character solutions adjoined with a constant character.

Furthermore, while performing the above classification, we also observe that though we began to search for solutions of Diophantine equations coming from $\mathbf{[3,0]}$ MLDE, we can obtain $\mathbf{[n,l]}$ solutions with $n\in\lbrace 1,2\rbrace$ and $l\in\lbrace 0,2,4\rbrace$. So, such solutions can be realised as solutions to $\mathbf{[3,0]}$ MLDE with the additional character(s) being either unstable or have negative $q$-expansion coefficients. These $\mathbf{[n,l]}$ solutions have been previously discussed in this MLDE approach to classification of RCFTs in the context of $1$-character and $2$-character RCFTs. To our surprise we find that all such $\mathbf{[n,l]}$ solutions could be realised as a solutions to the $\mathbf{[3,0]}$ MLDE. We can, in future, investigate the case of $\mathbf{[2,6]}$ and $\mathbf{[2,8]}$ theories as solutions to some higher-order MLDE inspired by our above observations.

In this paper, we have directly solved the MLDEs. In the process we have uncovered many phenomena. (i) It is possible for an infinite number of CFTs to all have the same central charge and scaling dimensions: sections \ref{23ss} and  \ref{313ss} (ii) It is possible for a CFT to have two identical characters: section \ref{312ss} (iii) A $\mathbf{[n,l]}$
CFT can occur as a solution to $\mathbf{[N,L]}$ CFT for $\mathbf{n} < \mathbf{N}$: section \ref{312ss} (iv) One can adjoin a constant character to a $\mathbf{[n,l]}$ CFT and obtain a $\mathbf{[n+1,n+l]}$ CFT: section \ref{32ss}. We have seen all these phenomena in the context of  $\mathbf{[3,0]}$ and  $\mathbf{[3,2]}$ CFTs and expect them to occur for higher characters as well. 

We have established a basic template on how to go about solving directly a MLDE: expressing the parameters in terms of objects associated with the identity character, seeking out Diophantine equations that exist because of the integrality of the Fourier coefficients of the identity character, making the problem 	``finite'' etc. We hope to bring this template to bear upon other $\mathbf{[n,l]}$ CFTs.  

Our observations in section \ref{6ss} suggest that there are more examples of the novel coset CFTs of Gaberdiel-Hampapura-Mukhi, which we hope to understand  \cite{wp}.  Admissible character solutions in category $\mathbf{III}$ and $\mathbf{V}$ which are to be considered still unknown, which are also potentially GHM-dual to known WZW CFTs, have better prospects of being consistent CFTs. We hope to report on these ongoing studies. 
	
\textbf{Note Added:} During the final stages of this work we became aware of concurrent work on this subject. The first paper \cite{Kaidi:2021ent} appeared while the first version of our paper was being prepared. The second paper \cite{Bae:2021mej} appeared very soon after the first version of our paper and before the present second version of our paper. There is an overlap of our $\mathbf{[3,0]}$ results with (small) parts of both these papers but  the methods seem to be quite different.

\begin{center}
\textbf{Acknowledgments}
\end{center}
AD would like to express his gratitude to the School of Physics, NISER, Bhubaneswar for the hospitality and resources required to complete this project. AD would like to thank Sigma Samhita for her immense help in the type setting of the tables required for this work. AD would also like to thank Jishu Das for useful discussions on modular forms. AD would also like to thank Rinkesh Panigrahi for discussions on Python. CNG thanks Jaban Meher for helpful discussions and Avani for lending him the use of her computer. JS is indebted to Sumedha and Senthil K. Kumar for useful discussions.

\clearpage
\appendix

\section{Tabulating the character-like solutions of the $\mathbf{[3,0]}$ and $\mathbf{[3,2]}$ MLDEs} \label{app1}


\begin{table}[!hbtp] 
\begin{center}
\begin{threeparttable}
\caption{Solutions to the $\mathbf{[3,0]}$ MLDE with $\mathbf{0<c\leq 7}$}\label{t7}
\begin{tabular*}{\textwidth}{@{\extracolsep{\fill}}cccc||cc||ccc||c}
\hline
\hline
\makebox[0pt][l]{\fboxsep0pt\colorbox{Mywhite} {\strut\hspace*{\linewidth}}}
$N$ &  $m_1$ & $m_2$ & $k$ & $D_1$ & $D_2$ & $c$ & $h_1$ & $h_2$ & \textbf{Category}  \\
\hline
\hline
\makebox[0pt][l]{\fboxsep0pt\colorbox{Mywhite} {\strut\hspace*{\linewidth}}}
35  & 0 & 1 & 265825 & $\cdots$ & $\cdots$ & 1/2  & 1/16 & 1/2  & {\bf I} \\
\makebox[0pt][l]{\fboxsep0pt\colorbox{Mygrey} {\strut\hspace*{\linewidth}}}
40 & 1 & 2 & 187200 & $\cdots$ & $\cdots$ &  4/7 & 1/7 & 3/7 & {\bf I}  \\
\makebox[0pt][l]{\fboxsep0pt\colorbox{Mywhite} {\strut\hspace*{\linewidth}}}
56 &  2 & 3 & 178752 & $\cdots$ & $\cdots$ & 4/5 & 1/5 & 2/5 & {\bf I} \\
\makebox[0pt][l]{\fboxsep0pt\colorbox{Mygrey} {\strut\hspace*{\linewidth}}}
70 &  1 & 4 & 441000 & 1 & 1 & 1 & 1/8 & 1/2 & {\bf I}  \\
\makebox[0pt][l]{\fboxsep0pt\colorbox{Mywhite} {\strut\hspace*{\linewidth}}}
70 &  3 & 4 & 137200 & $\cdots$ & $\cdots$ & 1 & 1/4 & 3/8 & {\bf II}  \\
\hline
\hline
\makebox[0pt][l]{\fboxsep0pt\colorbox{Mygrey} {\strut\hspace*{\linewidth}}}
105 &  3 & 9 & 532875 & $\cdots$ & $\cdots$ & 3/2 & 3/16 & 1/2 & {\bf I} \\
\makebox[0pt][l]{\fboxsep0pt\colorbox{Mywhite} {\strut\hspace*{\linewidth}}}
120 &  6 & 12 & 264000 & 3 & 2 & 12/7 & 2/7 & 3/7 & {\bf III} \\
\makebox[0pt][l]{\fboxsep0pt\colorbox{Mygrey} {\strut\hspace*{\linewidth}}}
140 &  6 & 17 & 548800 & $\cdots$ & $\cdots$ & 2 & 1/4 & 1/2 & {\bf I} \\
\makebox[0pt][l]{\fboxsep0pt\colorbox{Mywhite} {\strut\hspace*{\linewidth}}}
140 &  8 & 17 & 176400 & $\cdots$ & $\cdots$ & 2 & 1/3 & 5/12 & {\bf II} \\
\hline
\hline
\makebox[0pt][l]{\fboxsep0pt\colorbox{Mygrey} {\strut\hspace*{\linewidth}}}
168 &  3 & 18 & 1119552 & 3 & 1 & 12/5 & 1/5 & 3/5 & {\bf III} \\
\makebox[0pt][l]{\fboxsep0pt\colorbox{Mywhite} {\strut\hspace*{\linewidth}}}
175 &  10 & 30 & 496125 & $\cdots$ & $\cdots$ & 5/2 & 5/16 & 1/2 & {\bf I} \\
\makebox[0pt][l]{\fboxsep0pt\colorbox{Mygrey} {\strut\hspace*{\linewidth}}}
196 &  14 & 42 & 142688 & $\cdots$ & $\cdots$ &  14/5 & 2/5 & 9/20 & {\bf II} \\
\makebox[0pt][l]{\fboxsep0pt\colorbox{Mywhite} {\strut\hspace*{\linewidth}}}
210 &  15 & 51 & 382200 & $\cdots$ & $\cdots$ & 3 & 3/8 & 1/2 & {\bf I} \\
\hline
\hline
\makebox[0pt][l]{\fboxsep0pt\colorbox{Mygrey} {\strut\hspace*{\linewidth}}}
245 &  21 & 84 & 214375 & $\cdots$ & $\cdots$ & 7/2 & 7/16 & 1/2 & {\bf I} \\
\makebox[0pt][l]{\fboxsep0pt\colorbox{Mywhite} {\strut\hspace*{\linewidth}}}
280 & 16 & 98 & 1411200 & $\cdots$ & $\cdots$ & 4 & 1/3 & 2/3 & {\bf I} \\
\makebox[0pt][l]{\fboxsep0pt\colorbox{Mygrey} {\strut\hspace*{\linewidth}}}
280 &  24 & 124 & 784000 & $\cdots$ & $\cdots$ & 4 & 2/5 & 3/5 & {\bf I} \\
\makebox[0pt][l]{\fboxsep0pt\colorbox{Mywhite} {\strut\hspace*{\linewidth}}}
280 &  28 & 134 & 0 & $\cdots$ & $\cdots$ &  4 & 1/2 & 1/2 & {\bf II} \\
\hline
\hline
\makebox[0pt][l]{\fboxsep0pt\colorbox{Mygrey} {\strut\hspace*{\linewidth}}}
315 &  36 & 207 & 253575 & $\cdots$ & $\cdots$ & 9/2 & 1/2 & 9/16 & {\bf I} \\
\makebox[0pt][l]{\fboxsep0pt\colorbox{Mywhite} {\strut\hspace*{\linewidth}}}
350 &  45 & 310 & 539000 & $\cdots$ & $\cdots$ & 5 & 1/2 & 5/8 & {\bf I} \\
\hline
\hline
\makebox[0pt][l]{\fboxsep0pt\colorbox{Mygrey} {\strut\hspace*{\linewidth}}}
364 &  52 & 377 & 214032 & $\cdots$ & $\cdots$ &  26/5 & 11/20 & 3/5 & {\bf II}  \\
\makebox[0pt][l]{\fboxsep0pt\colorbox{Mywhite} {\strut\hspace*{\linewidth}}}
385 &   55 & 451 & 848925 & $\cdots$ & $\cdots$ & 11/2 & 1/2 & 11/16 & {\bf I} \\
\makebox[0pt][l]{\fboxsep0pt\colorbox{Mygrey} {\strut\hspace*{\linewidth}}}
392 &  28 & 280 & 2283008 & $\cdots$ & $\cdots$ & 28/5 & 2/5 & 4/5 & {\bf I} \\
\makebox[0pt][l]{\fboxsep0pt\colorbox{Mywhite} {\strut\hspace*{\linewidth}}}
420 &  66 & 639 & 1176000 & $\cdots$ & $\cdots$ & 6 & 1/2 & 3/4 & {\bf I} \\
\makebox[0pt][l]{\fboxsep0pt\colorbox{Mygrey} {\strut\hspace*{\linewidth}}}
420 &  78 & 729 & 352800 & $\cdots$ & $\cdots$ & 6 & 7/12 & 2/3 & {\bf II}  \\
\hline
\hline
\makebox[0pt][l]{\fboxsep0pt\colorbox{Mywhite} {\strut\hspace*{\linewidth}}}
440 &  88 & 902 & 598400 & 11 & 44 & 44/7 & 4/7 & 5/7 & {\bf III}  \\
\makebox[0pt][l]{\fboxsep0pt\colorbox{Mygrey} {\strut\hspace*{\linewidth}}}
455 &  78 & 884 & 1512875 & $\cdots$ & $\cdots$ & 13/2 & 1/2 & 13/16 & {\bf I} \\
\makebox[0pt][l]{\fboxsep0pt\colorbox{Mywhite} {\strut\hspace*{\linewidth}}}
490 &  91 & 1197 & 1852200 & $\cdots$ & $\cdots$ & 7 & 1/2 & 7/8 & {\bf I} \\
\makebox[0pt][l]{\fboxsep0pt\colorbox{Mygrey} {\strut\hspace*{\linewidth}}}
490 &  133 & 1673 & 411600 & $\cdots$ & $\cdots$ & 7 & 5/8 & 3/4 & {\bf II} \\
\hline
\hline
\end{tabular*} 
\end{threeparttable}
\end{center}
\end{table}

\begin{table}[h] 
\begin{center}
\begin{threeparttable}
\caption{Solutions to the $\mathbf{[3,0]}$ MLDE with $\mathbf{7<c\leq 12}$}\label{t8}
\begin{tabular*}{\textwidth}{@{\extracolsep{\fill}}cccc||cc||ccc||c}
\hline
\hline
\makebox[0pt][l]{\fboxsep0pt\colorbox{Mywhite} {\strut\hspace*{\linewidth}}}
$N$ &  $m_1$ & $m_2$ & $k$ & $D_1$ & $D_2$ &$c$ & $h_1$ & $h_2$ &  \textbf{Category}  \\
\hline
\hline
\makebox[0pt][l]{\fboxsep0pt\colorbox{Mywhite} {\strut\hspace*{\linewidth}}}
504 &  144 & 1926 & 620928 & 12 & 9 &  36/5 & 3/5 & 4/5 & {\bf III}  \\
\makebox[0pt][l]{\fboxsep0pt\colorbox{Mygrey} {\strut\hspace*{\linewidth}}}
520 &  156 & 2236 & 832000 & 13 & 78 & 52/7 & 4/7 & 6/7 & {\bf III} \\
\makebox[0pt][l]{\fboxsep0pt\colorbox{Mywhite} {\strut\hspace*{\linewidth}}}
525 &  105 & 1590 & 2186625 & $\cdots$ & $\cdots$ & 15/2 & 1/2 & 15/16 & {\bf I} \\
\makebox[0pt][l]{\fboxsep0pt\colorbox{Mygrey} {\strut\hspace*{\linewidth}}}
532 &  190 & 2831 & 268128 & $\cdots$ & $\cdots$ & 38/5 & 13/20 & 4/5 & {\bf II} \\
\makebox[0pt][l]{\fboxsep0pt\colorbox{Mywhite} {\strut\hspace*{\linewidth}}}
560 &  56 & 1052 & 3763200 & $\cdots$ & $\cdots$ & 8 & 1/2 & 1 & {\bf I} \\
\makebox[0pt][l]{\fboxsep0pt\colorbox{Mygrey} {\strut\hspace*{\linewidth}}}
560 &  120 & 2076 & 2508800 & $\cdots$ & $\cdots$ & 8 & 1/2 & 1 & {\bf I} \\
\makebox[0pt][l]{\fboxsep0pt\colorbox{Mywhite} {\strut\hspace*{\linewidth}}}
560 &  248 & 4124 & 0 & $\cdots$ & $\cdots$ & 8 & 3/4 & 3/4 & {\bf II} \\
\makebox[0pt][l]{\fboxsep0pt\colorbox{Mygrey} {\strut\hspace*{\linewidth}}}
560 &  $d_1$ & $d_2$ & $k_1$ & 1 & 1 &  8 & 1/2 & 1 & {\bf III} \\
\makebox[0pt][l]{\fboxsep0pt\colorbox{Mywhite} {\strut\hspace*{\linewidth}}}
560 &  $e_1$ & $e_2$ & $k_2$ & 1 & 1 & 8 & 1/2 & 1 & {\bf III} \\
\hline
\hline
\makebox[0pt][l]{\fboxsep0pt\colorbox{Mygrey} {\strut\hspace*{\linewidth}}}
595 &136 & 2669 & 2811375 & $\cdots$ & $\cdots$ &  17/2 & 1/2 & 17/16 &  {\bf I}  \\
\makebox[0pt][l]{\fboxsep0pt\colorbox{Mywhite} {\strut\hspace*{\linewidth}}}
595 &  255 & 4216 & 478975 & 17 & 221 & 17/2 & 1/16 & 3/2 & {\bf III} \\
\makebox[0pt][l]{\fboxsep0pt\colorbox{Mygrey} {\strut\hspace*{\linewidth}}}
600 & 210 & 4050 & 1560000 & 10 & 285 &  60/7 & 3/7 & 8/7 & {\bf III} \\
\makebox[0pt][l]{\fboxsep0pt\colorbox{Mywhite} {\strut\hspace*{\linewidth}}}
616 & 220 & 4433 & 1655808 & 11 & 11 &  44/5 & 2/5 & 6/5 &  {\bf III} \\
\makebox[0pt][l]{\fboxsep0pt\colorbox{Mygrey} {\strut\hspace*{\linewidth}}}
616 &  253 & 4642 & 931392 & 11 & 242 &44/5 & 1/5 & 7/5 & {\bf III} \\
\makebox[0pt][l]{\fboxsep0pt\colorbox{Mywhite} {\strut\hspace*{\linewidth}}}
630 & 153 & 3384 & 3087000 & $\cdots$ & $\cdots$ & 9 & 1/2 & 9/8 &  {\bf I} \\
\makebox[0pt][l]{\fboxsep0pt\colorbox{Mygrey} {\strut\hspace*{\linewidth}}}
630 &  261 & 4500 & 970200 & 9 & 57 & 9 & 1/8 & 3/2 & {\bf III} \\
\hline
\hline
\makebox[0pt][l]{\fboxsep0pt\colorbox{Mywhite} {\strut\hspace*{\linewidth}}}
665 &  171 & 4237 & 3328325 & $\cdots$ & $\cdots$ & 19/2 & 1/2 & 19/16 & {\bf I}  \\
\makebox[0pt][l]{\fboxsep0pt\colorbox{Mygrey} {\strut\hspace*{\linewidth}}}
665 & 266 & 4997 & 1466325 & 19 & 703 & 19/2 & 3/16 & 3/2 &  {\bf III} \\
\makebox[0pt][l]{\fboxsep0pt\colorbox{Mywhite} {\strut\hspace*{\linewidth}}}
680 &  221 & 5474 & 2692800 & 17 & 782 & 68/7 & 3/7 & 9/7 & {\bf III} \\
\makebox[0pt][l]{\fboxsep0pt\colorbox{Mygrey} {\strut\hspace*{\linewidth}}}
700 & 190 & 5245 & 3528000 & $\cdots$ & $\cdots$ & 10 & 1/2 & 5/4 &  {\bf I}  \\
\makebox[0pt][l]{\fboxsep0pt\colorbox{Mywhite} {\strut\hspace*{\linewidth}}}
700 &  270 & 5725 & 1960000 & 5 & 15 & 10 & 1/4 & 3/2 & {\bf III} \\
\hline
\hline
\makebox[0pt][l]{\fboxsep0pt\colorbox{Mygrey} {\strut\hspace*{\linewidth}}}
728 &  104 & 3458 & 5136768 & $\cdots$ & $\cdots$ &52/5 & 3/5 & 6/5 & {\bf I}  \\
\makebox[0pt][l]{\fboxsep0pt\colorbox{Mywhite} {\strut\hspace*{\linewidth}}}
735 &  210 & 6426 & 3678675 & $\cdots$ & $\cdots$ & 21/2 & 1/2 & 21/16 & {\bf I} \\
\makebox[0pt][l]{\fboxsep0pt\colorbox{Mygrey} {\strut\hspace*{\linewidth}}}
735 & 273 & 6699 & 2443875 & 21 & 1225 & 21/2 & 5/16 & 3/2 &  {\bf III} \\
\makebox[0pt][l]{\fboxsep0pt\colorbox{Mywhite} {\strut\hspace*{\linewidth}}}
770 & 231 & 7799 & 3773000 & $\cdots$ & $\cdots$ & 11 & 1/2 & 11/8 &  {\bf I} \\
\makebox[0pt][l]{\fboxsep0pt\colorbox{Mygrey} {\strut\hspace*{\linewidth}}}
770 &   275 & 7931 & 2910600 & 11 & 187 & 11 & 3/8 & 3/2 & {\bf III} \\
\hline
\hline
\makebox[0pt][l]{\fboxsep0pt\colorbox{Mywhite} {\strut\hspace*{\linewidth}}}
805 & 253 & 9384 & 3803625 & $\cdots$ & $\cdots$ & 23/2 & 1/2 & 23/16 &  {\bf I} \\
\makebox[0pt][l]{\fboxsep0pt\colorbox{Mygrey} {\strut\hspace*{\linewidth}}}
805 &  276 & 9430 & 3352825 & 23 & 1771 & 23/2 & 7/16 & 3/2 & {\bf III} \\
\makebox[0pt][l]{\fboxsep0pt\colorbox{Mywhite} {\strut\hspace*{\linewidth}}}
840 &  156 & 7542 & 5644800 & $\cdots$ & $\cdots$ &  12 & 2/3 & 4/3 & {\bf I} \\
\makebox[0pt][l]{\fboxsep0pt\colorbox{Mygrey} {\strut\hspace*{\linewidth}}}
840 &  222 & 10050 & 4704000 & 1 & 51 & 12 & 3/5 & 7/5 & {\bf III} \\
\makebox[0pt][l]{\fboxsep0pt\colorbox{Mywhite} {\strut\hspace*{\linewidth}}}
840 &  276 & 11202 & 3763200 & $\cdots$ & $\cdots$ & 12 & 1/2 & 3/2 & {\bf I} \\
\makebox[0pt][l]{\fboxsep0pt\colorbox{Mygrey} {\strut\hspace*{\linewidth}}}
840 &  318 & 8514 & 2822400 & 1 & 1 & 12 & 1/3 & 5/3 & {\bf V} \\
\hline
\hline
\end{tabular*} 
\end{threeparttable}
\end{center}
\end{table}

\begin{table}[h] 
\begin{center}
\begin{threeparttable}
\caption{Solutions to the $\mathbf{[3,0]}$ MLDE with $\mathbf{12<c\leq 17}$}\label{t9}
\begin{tabular*}{\textwidth}{@{\extracolsep{\fill}}cccc||cc||ccc||c}
\hline
\hline
\makebox[0pt][l]{\fboxsep0pt\colorbox{Mywhite} {\strut\hspace*{\linewidth}}}
$N$ &  $m_1$ & $m_2$ & $k$ & $D_1$ & $D_2$ & $c$ & $h_1$ & $h_2$ & \textbf{Category}  \\
\hline
\hline
\makebox[0pt][l]{\fboxsep0pt\colorbox{Mywhite} {\strut\hspace*{\linewidth}}}
875 &  275 & 13250 & 4134375 & 125 & 2325 & 25/2 & 9/16 & 3/2 & {\bf III}  \\
\makebox[0pt][l]{\fboxsep0pt\colorbox{Mygrey} {\strut\hspace*{\linewidth}}}
875 &  300 & 13275 & 3644375 & $\cdots$ & $\cdots$ & 25/2 & 1/2 & 25/16 & {\bf I} \\
\makebox[0pt][l]{\fboxsep0pt\colorbox{Mywhite} {\strut\hspace*{\linewidth}}}
910 & 273 & 15574 & 4459000 & 13 & 325 & 13 & 5/8 & 3/2 &  {\bf III} \\
\makebox[0pt][l]{\fboxsep0pt\colorbox{Mygrey} {\strut\hspace*{\linewidth}}}
910 &  325 & 15626 & 3439800 & $\cdots$ & $\cdots$ & 13 & 1/2 & 13/8 & {\bf I} \\
\hline
\hline
\makebox[0pt][l]{\fboxsep0pt\colorbox{Mywhite} {\strut\hspace*{\linewidth}}}
945 &  270 & 18171 & 4729725 & 27 & 2871 & 27/2 & 11/16 & 3/2 & {\bf III}  \\
\makebox[0pt][l]{\fboxsep0pt\colorbox{Mygrey} {\strut\hspace*{\linewidth}}}
945 &  351 & 18279 & 3142125 & $\cdots$ & $\cdots$ & 27/2 & 1/2 & 27/16 & {\bf I} \\
\makebox[0pt][l]{\fboxsep0pt\colorbox{Mywhite} {\strut\hspace*{\linewidth}}}
952 &  136 & 10438 & 6717312 & 119 & 68 & 68/5 & 4/5 & 7/5 & {\bf III} \\
\makebox[0pt][l]{\fboxsep0pt\colorbox{Mygrey} {\strut\hspace*{\linewidth}}}
952 &  374 & 11985 & 2612288 & 119 & 12138 & 68/5 & 2/5 & 9/5 & {\bf III} \\
\makebox[0pt][l]{\fboxsep0pt\colorbox{Mywhite} {\strut\hspace*{\linewidth}}}
980 &  266 & 21035 & 4939200 & $\cdots$ & $\cdots$ & 14 & 3/4 & 3/2 & {\bf I} \\
\makebox[0pt][l]{\fboxsep0pt\colorbox{Mygrey} {\strut\hspace*{\linewidth}}}
980 & 378 & 21259 & 2744000 & $\cdots$ & $\cdots$ & 14 & 1/2 & 7/4 &  {\bf I} \\
\hline
\hline
\makebox[0pt][l]{\fboxsep0pt\colorbox{Mywhite} {\strut\hspace*{\linewidth}}}
1000 & 325 & 26850 & 3960000 & 55 & 2925 & 100/7 & 5/7 & 11/7 & {\bf III}  \\
\makebox[0pt][l]{\fboxsep0pt\colorbox{Mygrey} {\strut\hspace*{\linewidth}}}
1000 & 380 & 26850 & 2816000 & 55 & 11495 & 100/7 & 4/7 & 12/7 &  {\bf III}  \\
\makebox[0pt][l]{\fboxsep0pt\colorbox{Mywhite} {\strut\hspace*{\linewidth}}}
1015 &  261 & 24157 & 5080075 & 29 & 3393 & 29/2 & 13/16 & 3/2 & {\bf III}  \\
\makebox[0pt][l]{\fboxsep0pt\colorbox{Mygrey} {\strut\hspace*{\linewidth}}}
1015 &  406 & 24592 & 2238075 & $\cdots$ & $\cdots$ & 29/2 & 1/2 & 29/16 & {\bf I} \\
\makebox[0pt][l]{\fboxsep0pt\colorbox{Mywhite} {\strut\hspace*{\linewidth}}}
1050 &  255 & 27525 & 5145000 & 15 & 455 & 15 & 7/8 & 3/2 & {\bf IV} \\
\makebox[0pt][l]{\fboxsep0pt\colorbox{Mygrey} {\strut\hspace*{\linewidth}}}
1050 &  435 & 28305 & 1617000 & $\cdots$ & $\cdots$ & 15 & 1/2 & 15/8 & {\bf I} \\
\hline
\hline
\makebox[0pt][l]{\fboxsep0pt\colorbox{Mywhite} {\strut\hspace*{\linewidth}}}
1064 & 380 & 41762 & 2860032 & 57 & 3249 & 76/5 & 4/5 & 8/5 & {\bf I}  \\
\makebox[0pt][l]{\fboxsep0pt\colorbox{Mygrey} {\strut\hspace*{\linewidth}}}
1064 &437 & 41762 & 1608768 & 57 & 19 &  76/5 & 3/5 & 9/5 &  {\bf III}  \\
\makebox[0pt][l]{\fboxsep0pt\colorbox{Mywhite} {\strut\hspace*{\linewidth}}}
1080 &  378 & 45009 & 2808000 & 117 & 3510 & 108/7 & 6/7 & 11/7 & {\bf III}  \\
\makebox[0pt][l]{\fboxsep0pt\colorbox{Mygrey} {\strut\hspace*{\linewidth}}}
1080 &  456 & 45126 & 1123200 & 39 & 20424 & 108/7 & 4/7 & 13/7 & {\bf III}  \\
\makebox[0pt][l]{\fboxsep0pt\colorbox{Mywhite} {\strut\hspace*{\linewidth}}}
1085 &  248 & 31124 & 5126625 & $\cdots$ & $\cdots$ & 31/2 & 15/16 & 3/2 & {\bf I} \\
\makebox[0pt][l]{\fboxsep0pt\colorbox{Mygrey} {\strut\hspace*{\linewidth}}}
1085 &  465 & 32426 & 873425 & $\cdots$ & $\cdots$ & 31/2 & 1/2 & 31/16 & {\bf I} \\
\makebox[0pt][l]{\fboxsep0pt\colorbox{Mywhite} {\strut\hspace*{\linewidth}}}
1120 &  496 & 36984 & 0 & $\cdots$ & $\cdots$ & 16 & 1/2 & 2 & {\bf I} \\
\makebox[0pt][l]{\fboxsep0pt\colorbox{Mygrey} {\strut\hspace*{\linewidth}}}
1120 &  496 & 69752 & 0 & $\cdots$ & $\cdots$ & 16 & 5/4 & 5/4 & {\bf II} \\
\makebox[0pt][l]{\fboxsep0pt\colorbox{Mywhite} {\strut\hspace*{\linewidth}}}
1120 & $d_1$ & $d_2$ & $k_1$ & 1 & 1 & 16 & 1 & 3/2 &  {\bf III} \\
\makebox[0pt][l]{\fboxsep0pt\colorbox{Mygrey} {\strut\hspace*{\linewidth}}}
1120 & $e_1$ & $e_2$ & $k_2$ & 1 & 1 &  16 & 1 & 3/2 &  {\bf III} \\
\hline
\hline
\makebox[0pt][l]{\fboxsep0pt\colorbox{Mywhite} {\strut\hspace*{\linewidth}}}
1148 &  410 & 64739 & 578592 & 902 & $\cdots$ &  82/5 & 6/5 & 27/20 & {\bf II} \\
\makebox[0pt][l]{\fboxsep0pt\colorbox{Mygrey} {\strut\hspace*{\linewidth}}}
1155 &  231 & 38940 & 4810575 & 33 & 4301 & 33/2 & 17/16 & 3/2 & {\bf III} \\
\makebox[0pt][l]{\fboxsep0pt\colorbox{Mywhite} {\strut\hspace*{\linewidth}}}
1155 &  528 & 42009 & 1010625 & $\cdots$ & $\cdots$ & 33/2 & 1/2 & 33/16 & {\bf I} \\
\makebox[0pt][l]{\fboxsep0pt\colorbox{Mygrey} {\strut\hspace*{\linewidth}}}
1160 &  348 & 58174 & 1856000 & 725 & 1972 & 116/7 & 8/7 & 10/7 & {\bf III}  \\
\makebox[0pt][l]{\fboxsep0pt\colorbox{Mywhite} {\strut\hspace*{\linewidth}}}
1176 &  336 & 59850 & 1448832 & 154 & 1452 & 84/5 & 6/5 & 7/5 & {\bf III} \\
\makebox[0pt][l]{\fboxsep0pt\colorbox{Mygrey} {\strut\hspace*{\linewidth}}}
1176 &  534 & 73842 & 1138368 & 33 & 55924 & 84/5 & 1/5 & 12/5 & {\bf III} \\
\makebox[0pt][l]{\fboxsep0pt\colorbox{Mywhite} {\strut\hspace*{\linewidth}}}
1190 &  221 & 43112 & 4498200 & 17 & 561 & 17 & 9/8 & 3/2 & {\bf IV} \\
\makebox[0pt][l]{\fboxsep0pt\colorbox{Mygrey} {\strut\hspace*{\linewidth}}}
1190 &  323 & 60860 & 999600 & 51 & $\cdots$ & 17 & 5/4 & 11/8 & {\bf II} \\
\makebox[0pt][l]{\fboxsep0pt\colorbox{Mywhite} {\strut\hspace*{\linewidth}}}
1190 &  561 & 47532 & 2165800 & $\cdots$ & $\cdots$ & 17 & 1/2 & 17/8 & {\bf I} \\
\hline
\hline
\end{tabular*} 
\end{threeparttable}
\end{center}
\end{table}

\begin{table}[h] 
\begin{center}
\begin{threeparttable}
\caption{Solutions to the $\mathbf{[3,0]}$ MLDE with $\mathbf{17<c\leq 21}$}\label{t10}
\rowcolors{2}{Mygrey}{Mywhite}
\begin{tabular}{cccc||cc||ccc||c}
\hline
\hline
$N$ &  $m_1$ & $m_2$ & $k$ & $D_1$ & $D_2$ & $c$ & $h_1$ & $h_2$ & \textbf{Category}  \\
\hline
\hline
1225 &  210 & 47425 & 4073125 & 35 & 4655 & 35/2 & 19/16 & 3/2 & {\bf IV} \\
1225 &  595 & 53585 & 3472875 & $\cdots$ & $\cdots$ & 35/2 & 1/2 & 35/16 & {\bf I} \\
1240 &  248 & 58212 & 1686400 & 2108 & 2108 & 124/7 & 9/7 & 10/7 & {\bf III}  \\
1260 &  198 & 51849 & 3528000 & 9 & 75 & 18 & 5/4 & 3/2 & {\bf IV} \\
1260 &  234 & 59805 & 1058400 & 1 & $\cdots$ & 18 & 4/3 & 17/12 & {\bf II}  \\
1260 &  598 & 83049 & 3528000 & 25 & 221 & 18 & 1/4 & 5/2 & {\bf III}  \\
1260 &  630 & 60201 & 4939200 & $\cdots$ & $\cdots$ & 18 & 1/2 & 9/4 & {\bf I}  \\
\hline
\hline
1288 & 92 & 30774 & 7501312 & 1196 & 299 & 92/5 & 6/5 & 8/5 &  {\bf III} \\
1288 &  690 & 36754 & 7501312 & 299 & 178802 & 92/5 & 3/5 & 11/5 & {\bf III} \\
1295 &  185 & 56351 & 2855475 & 37 & 4921 & 37/2 & 21/16 & 3/2 & {\bf IV}  \\
1295 &  666 & 67414 & 6572125 & $\cdots$ & $\cdots$ & 37/2 & 1/2 & 37/16 &  {\bf I}  \\
1316 &  188 & 62087 & 773808 & 4794 & $\cdots$ & 94/5 & 7/5 & 29/20 & {\bf II} \\
1330 &  171 & 60895 & 2048200 & 19 & 627 & 19 & 11/8 & 3/2 & {\bf IV}  \\
1330 &  703 & 75259 & 8379000 & $\cdots$ & $\cdots$ & 19 & 1/2 & 19/8 & {\bf I}  \\
\hline
\hline
1365 &156 & 65442 & 1098825 & 39 & 5083 &  39/2 & 23/16 & 3/2 &  {\bf IV} \\
1365 &  741 & 83772 & 10367175 & $\cdots$ & $\cdots$ & 39/2 & 1/2 & 39/16 & {\bf I} \\
1400 &  80 & 46790 & 7056000 & 5 & 4 & 20 & 4/3 & 5/3 & {\bf V} \\
1400 &  120 & 62630 & 3920000 & 4 & 13 & 20 & 7/5 & 8/5 & {\bf IV}  \\
1400 &  140 & 69950 & 0 & 5 & 5 & 20 & 3/2 & 3/2 & {\bf II} \\
1400 &  728 & 106406 & 9878400 & 2 & 1 &  20 & 1/3 & 8/3 & {\bf V}  \\
1400 &  780 & 92990 & 12544000 & $\cdots$ & $\cdots$ & 20 & 1/2 & 5/2 & {\bf I}  \\
1400 &  890 & 55700 & 17640000 & 5 & 10 & 20 & 2/3 & 7/3 & {\bf V}  \\
\hline
\hline
1435 & 123 & 74374 & 1255625 & 5125 & 41 & 41/2 & 3/2 & 25/16 &  {\bf IV} \\
1435 &  820 & 102951 & 14916825 & $\cdots$ & $\cdots$ & 41/2 & 1/2 & 41/16 & {\bf I} \\
1470 &  105 & 78666 & 2675400 & 637 & 21 & 21 & 3/2 & 13/8 & {\bf IV}  \\
1470 &  861 & 113694 & 17493000 & $\cdots$ & $\cdots$ & 21 & 1/2 & 21/8 & {\bf I}  \\
\hline
\hline
\end{tabular} 
\end{threeparttable}
\end{center}
\end{table}

\begin{table}[h] 
\begin{center}
\begin{threeparttable}
\caption{Solutions to the $\mathbf{[3,0]}$ MLDE with $\mathbf{21<c\leq 26}$}\label{t11}
\rowcolors{2}{Mygrey}{Mywhite}
\begin{tabular}{cccc||cc||ccc||c}
\hline
\hline
$N$ &  $m_1$ & $m_2$ & $k$ & $D_1$ & $D_2$ & $c$ & $h_1$ & $h_2$ & \textbf{Category}  \\
\hline
\hline
1484 &  106 & 84429 & 1080352 & $\cdots$ & 15847 & 106/5 & 31/20 & 8/5 & {\bf II} \\
1505 & 86 & 82775 & 4266675 & 5031 & 43 & 43/2 & 3/2 & 27/16 &  {\bf IV} \\
1505 &  903 & 125259 & 20279875 & $\cdots$ & $\cdots$ &  43/2 & 1/2 & 43/16 & {\bf I} \\
1512 &  27 & 46386 & 10075968 & 459 & 42483 & 108/5 & 7/5 & 9/5 & {\bf III} \\
1512 &  860 & 133851 & 17912832 & 833 & 3015426 & 108/5 & 2/5 & 14/5 & {\bf III} \\
1512 &  1404 & 53730 & 46061568 & 459 & 153 & 108/5 & 4/5 & 12/5 & {\bf III} \\
1540 &  66 & 86647 & 6036800 & 77 & 11 & 22 & 3/2 & 7/4 & {\bf III} \\
1540 &  88 & 99935 & 1940400 & $\cdots$ & 22 & 22 & 19/12 & 5/3 & {\bf II} \\
1540 &  946 & 137687 & 23284800 & $\cdots$ & $\cdots$ & 22 & 1/2 & 11/4 & {\bf I}  \\
1540 &  1298 & 98967 & 42257600 & 77 & 847 & 22 & 3/4 & 5/2 & {\bf III} \\
\hline
\hline
1560 &  78 & 104754 & 3432000 & 5070 & 27170 & 156/7 & 11/7 & 12/7 & {\bf III} \\
1560 &  1248 & 120874 & 40560000 & 130 & 799500 & 156/7 & 5/7 & 18/7 & {\bf III} \\
1575 &  45 & 90225 & 7993125 & 4785 & 45 & 45/2 & 3/2 & 29/16 & {\bf IV} \\
1575 &  990 & 151020 & 26515125 & $\cdots$ & $\cdots$ & 45/2 & 1/2 & 45/16 & {\bf I} \\
1575 & 1640 & 99795 & 62346375 & 1595 & 956449 & 45/2 & 13/16 & 5/2 &  {\bf III} \\
1610 &  23 & 93449 & 10143000 & 575 & 23 & 23 & 3/2 & 15/8 & {\bf III} \\
1610 & 69 & 131905 & 3155600 & $\cdots$ & 253 & 23 & 13/8 & 7/4 & {\bf II} \\
1610 &  1035 & 165301 & 29978200 & $\cdots$ & $\cdots$ & 23 & 1/2 & 23/8 & {\bf I}  \\
1610 &  2323 & 100349 & 102557000 & 575 & 32683 & 23 & 7/8 & 5/2 & {\bf III} \\
\hline
\hline
1624 &  58 & 136706 & 5183808 & 4959 & 1102 & 116/5 & 8/5 & 9/5 & {\bf III} \\
1624 &  1711 & 146624 & 69981408 & 1653 & 910803 & 116/5 & 4/5 & 13/5 & {\bf III} \\
1640 &  41 & 141122 & 7675200 & 4797 & 50922 & 164/7 & 11/7 & 13/7 & {\bf III} \\
1645 &  0 & 96256 & 12493775 & 4371 & 47 & 47/2 & 3/2 & 31/16 & {\bf IV} \\
1645 &  1081 & 180574 & 33681375 & $\cdots$ & $\cdots$ & 47/2 & 1/2 & 47/16 &  {\bf I} \\
1645 & 4371 & 100627 & 223103125 & 4371 & 1135003 & 47/2 & 15/16 & 5/2 &  {\bf III} \\
1652 &  59 & 164315 & 3954888 & $\cdots$ & 32509 & 118/5 & 33/20 & 9/5 & {\bf II} \\
1680 & 1128 & 196884 & 37632000 & $\cdots$ & $\cdots$ & 24 & 1/2 & 3 &  {\bf I}  \\
\hline
\hline
1715 & 1176 & 214277 & 41837425 & $\cdots$ & $\cdots$ & 49/2 & 1/2 & 49/16 & {\bf I} \\
1750 &  1225 & 232800 & 46305000 & $\cdots$ & $\cdots$ & 25 & 1/2 & 25/8 & {\bf I}  \\
\hline
\hline
1785 &  1275 & 252501 & 51042075 & $\cdots$ & $\cdots$ & 51/2 & 1/2 & 51/16 & {\bf I} \\
1820 &  1118 & 258869 & 39748800 & 117 & 3315 & 26 & 1/4 & 7/2 & {\bf III}  \\
1820 &  1326 & 273429 & 56056000 & $\cdots$ & $\cdots$ &  26 & 1/2 & 13/4 & {\bf I}  \\
\hline
\hline
\end{tabular} 
\end{threeparttable}
\end{center}
\end{table}

\begin{table}[h] 
\begin{center}
\begin{threeparttable}
\caption{Solutions to the $\mathbf{[3,0]}$ MLDE with $\mathbf{26<c\leq 36}$}\label{t12}
\rowcolors{2}{Mygrey}{Mywhite}
\begin{tabular}{cccc||cc||ccc||c}
\hline
\hline
$N$ &  $m_1$ & $m_2$ & $k$ & $D_1$ & $D_2$ &$c$ & $h_1$ & $h_2$ & \textbf{Category}  \\
\hline
\hline
1848 &  1536 & 305286 & 73137792 & 2392 & 47018049 & 132/5 & 3/5 & 16/5 & {\bf III}  \\
1855 &  1378 & 295634 & 61354125 & $\cdots$ & $\cdots$ &  53/2 & 1/2 & 53/16 & {\bf I} \\
1890 & 1431 & 319167 & 66943800 & $\cdots$ & $\cdots$ & 27 & 1/2 & 27/8 &  {\bf I} \\
\hline
\hline
1925 &  1485 & 344080 & 72832375 & $\cdots$ & $\cdots$ & 55/2 & 1/2 & 55/16 & {\bf I} \\
1960 & 1540 & 370426 & 79027200 & $\cdots$ & $\cdots$ & 28 & 1/2 & 7/2 &  {\bf I} \\
1960 &  1948 & 424724 & 112896000 & 25 & 11 & 28 & 2/3 & 10/3 & {\bf V} \\
\hline
\hline
1995 &  1596 & 398259 & 85535625 & $\cdots$ & $\cdots$ & 57/2 & 1/2 & 57/16 & {\bf I} \\
2030 &  1653 & 427634 & 92365000 & $\cdots$ & $\cdots$ & 29 & 1/2 & 29/8 & {\bf I} \\
\hline
\hline
2065 &  1711 & 458607 & 99522675 & $\cdots$ & $\cdots$ & 59/2 & 1/2 & 59/16 & {\bf I} \\
2100 &  1770 & 491235 & 107016000 & $\cdots$ & $\cdots$ & 30 & 1/2 & 15/4 & {\bf I} \\
2100 &  2778 & 683715 & 199214400 & 539 & 14421 & 30 & 3/4 & 7/2 & {\bf III} \\
\hline
\hline
2135 &  1830 & 525576 & 114852325 & $\cdots$ & $\cdots$ & 61/2 & 1/2 & 61/16 & {\bf I} \\
2135 &  3599 & 888770 & 279546225 & 47763 & 264580485 & 61/2 & 13/16 & 7/2 & {\bf III} \\
2170 &  1891 & 561689 & 123039000 & $\cdots$ & $\cdots$ & 31 & 1/2 & 31/8 & {\bf I} \\
2170 &  5239 & 1296885 & 440206200 & 9269 & 2295147 & 31 & 7/8 & 7/2 & {\bf III} \\
\hline
\hline
2184 &  3612 & 958572 & 290293248 & 14877 & 250774426 & 156/5 & 4/5 & 18/5 & {\bf III} \\
2205 &  1953 & 599634 & 131583375 & $\cdots$ & $\cdots$ & 63/2 & 1/2 & 63/16 & {\bf I} \\
2240 &  2016 & 639472 & 140492800 & $\cdots$ & $\cdots$ &32 & 1/2 & 4 & {\bf I} \\
\hline
\hline
2268 &  4 & 113253 & 5882352 & 310124 & $\cdots$ & 162/5 & 11/5 & 47/20 & {\bf II} \\
2275 &  2080 & 681265 & 149774625 & $\cdots$ & $\cdots$ & 65/2 & 1/2 & 65/16 & {\bf I} \\
2296 &  0 & 90118 & 7971712 & 10168 & 615164 & 164/5 & 11/5 & 12/5 & {\bf IV} \\
2310 &  3 & 86004 & 4998000 & 1105 & $\cdots$ & 33 & 9/4 & 19/8 & {\bf II} \\
2310 &  2145 & 725076 & 159436200 & $\cdots$ & $\cdots$ & 33 & 1/2 & 33/8 & {\bf I} \\
\hline
\hline
2345 &  2211 & 770969 & 169484875 & $\cdots$ & $\cdots$ & 67/2 & 1/2 & 67/16 & {\bf I} \\
2360 &  0 & 63366 & 5852800 & 715139 & 848656 & 236/7 & 16/7 & 17/7 & {\bf IV} \\
2380 &  1 & 58997 & 3439800 & 13 & $\cdots$ & 34 & 7/3 & 29/12 & {\bf II} \\
2380 &  2278 & 819009 & 179928000 & $\cdots$ & $\cdots$ & 34 & 1/2 & 17/4 & {\bf I} \\
\hline
\hline
2415 & 2346 & 869262 & 190772925 & $\cdots$ & $\cdots$ &  69/2 & 1/2 & 69/16 & {\bf I} \\
2450 &  2415 & 921795 & 202027000 & $\cdots$ & $\cdots$ & 35 & 1/2 & 35/8 & {\bf I} \\
\hline
\hline
2485 &  2485 & 976676 & 213697575 & $\cdots$ & $\cdots$ & 71/2 & 1/2 & 71/16 & {\bf I} \\
2520 &  2556 & 1033974 & 225792000 & $\cdots$ & $\cdots$ & 36 & 1/2 & 9/2 & {\bf I} \\
2520 &  3384 & 1337850 & 325987200 & 2 & 4 & 36 & 2/3 & 13/3 & {\bf V} \\
\hline
\hline
\end{tabular} 
\end{threeparttable}
\end{center}
\end{table}

\begin{table}[h] 
\begin{center}
\begin{threeparttable}
\caption{Solutions to the $\mathbf{[3,0]}$ MLDE with $\mathbf{36<c\leq 51}$}\label{t13}
\rowcolors{2}{Mygrey}{Mywhite}
\begin{tabular}{cccc||cc||ccc||c}
\hline
\hline
$N$ &  $m_1$ & $m_2$ & $k$ & $D_1$ & $D_2$ & $c$ & $h_1$ & $h_2$ & \textbf{Category}  \\
\hline
\hline
2555 &  2628 & 1093759 & 238317625 & $\cdots$ & $\cdots$ & 73/2 & 1/2 & 73/16 & {\bf I} \\
2590 &  2701 & 1156102 & 251281800 & $\cdots$ & $\cdots$ & 37 & 1/2 & 37/8 & {\bf I} \\
\hline
\hline
2625 &  2775 & 1221075 & 264691875 & $\cdots$ & $\cdots$ & 75/2 & 1/2 & 75/16 & {\bf I} \\
2660 &  2850 & 1288751 & 278555200 & $\cdots$ & $\cdots$ & 38 & 1/2 & 19/4 & {\bf I} \\
\hline
\hline
2695 &  2926 & 1359204 & 292879125 & $\cdots$ & $\cdots$ & 77/2 & 1/2 & 77/16 & {\bf I} \\
2730 &  3003 & 1432509 & 307671000 & $\cdots$ & $\cdots$ & 39 & 1/2 & 39/8 & {\bf I} \\
\hline
\hline
2765 & 3081 & 1508742 & 322938175 & $\cdots$ & $\cdots$ &  79/2 & 1/2 & 79/16 & {\bf I} \\
2800 &  3160 & 1587980 & 338688000 & $\cdots$ & $\cdots$ & 40 & 1/2 & 5 & {\bf I} \\
\hline
\hline
2835 &  3240 & 1670301 & 354927825 & $\cdots$ & $\cdots$ & 81/2 & 1/2 & 81/16 & {\bf I} \\
2870 &  3321 & 1755784 & 371665000 & $\cdots$ & $\cdots$ & 41 & 1/2 & 41/8 & {\bf I} \\
\hline
\hline
2905 &  3403 & 1844509 & 388906875 & $\cdots$ & $\cdots$ & 83/2 & 1/2 & 83/16 & {\bf I} \\
2940 &  3486 & 1936557 & 406660800 & $\cdots$ & $\cdots$ & 42 & 1/2 & 21/4 & {\bf I} \\
\hline
\hline
2975 & 3570 & 2032010 & 424934125 & $\cdots$ & $\cdots$ &  85/2 & 1/2 & 85/16 &  {\bf I} \\
3010 &  3655 & 2130951 & 443734200 & $\cdots$ & $\cdots$ & 43 & 1/2 & 43/8 & {\bf I} \\
\hline
\hline
3045 &  3741 & 2233464 & 463068375 & $\cdots$ & $\cdots$ & 87/2 & 1/2 & 87/16 & {\bf I} \\
3080 &  3146 & 1906130 & 372556800 & 13 & 19 & 44 & 1/3 & 17/3 & {\bf V} \\
3080 &  3828 & 2339634 & 482944000 & $\cdots$ & $\cdots$ & 44 & 1/2 & 11/2 & {\bf I} \\
\hline
\hline
3115 &  3916 & 2449547 & 503368425 & $\cdots$ & $\cdots$ & 89/2 & 1/2 & 89/16 & {\bf I} \\
3150  & 4005 & 2563290 & 524349000 & $\cdots$ & $\cdots$ & 45 & 1/2 & 45/8 & {\bf I} \\
\hline
\hline
3185 & 4095 & 2680951 & 545893075 & $\cdots$ & $\cdots$ & 91/2 & 1/2 & 91/16 & {\bf I} \\
3220 &  4186 & 2802619 & 568008000 & $\cdots$ & $\cdots$ & 46 & 1/2 & 23/4 & {\bf I} \\
\hline
\hline
3255 &  4278 & 2928384 & 590701125 & $\cdots$ & $\cdots$ & 93/2 & 1/2 & 93/16 & {\bf I} \\
3290 & 4371 & 3058337 & 613979800 & $\cdots$ & $\cdots$ & 47 & 1/2 & 47/8 &  {\bf I} \\
\hline
\hline
3325 &  4465 & 3192570 & 637851375 & $\cdots$ & $\cdots$ & 95/2 & 1/2 & 95/16 & {\bf I} \\
3360 &  4560 & 3331176 & 662323200 & $\cdots$ & $\cdots$ & 48 & 1/2 & 6 & {\bf I} \\
\hline
\hline
$3395$ & $4656$ & $3474249$ & $687402625$ & $\cdots$ & $\cdots$ & 97/2 & 1/2 & 97/16 & {\bf I} \\
$3430$ & $4753$ & $3621884$ & $713097000$ & $\cdots$ & $\cdots$ & 49 & 1/2 & 49/8  & {\bf I} \\
\hline
\hline
$3465$ & $4851$ & $3774177$ & $739413675$ & $\cdots$ & $\cdots$ & 99/2 & 1/2 & 99/16 & {\bf I} \\
$3500$ & $4950$ & $3931225$ & $766360000$ & $\cdots$ & $\cdots$ & 50 & 1/2 & 25/4  & {\bf I} \\
\hline
\hline
$3535$ & $5050$ & $4093126$ & $793943325$ & $\cdots$ & $\cdots$ & 101/2 & 1/2 & 101/16 & {\bf I} \\
$3570$ & $5151$ & $4259979$ & $822171000$ &  $\cdots$ & $\cdots$ &  51 & 1/2 & 51/8  &
 {\bf I} \\
\hline
\hline
\end{tabular} 
\end{threeparttable}
\end{center}
\end{table}

\begin{table}[h] 
\begin{center}
\begin{threeparttable}
\caption{Solutions to the $\mathbf{[3,0]}$ MLDE with $\mathbf{51<c\leq 56}$}\label{t14}
\rowcolors{2}{Mygrey}{Mywhite}
\begin{tabular}{cccc||cc||ccc||c}
\hline
\hline
$N$ &  $m_1$ & $m_2$ & $k$ & $D_1$ & $D_2$ & $c$ & $h_1$ & $h_2$ & \textbf{Category}  \\
\hline
\hline
$3605$ & $5253$ & $4431884$ & $851050375$ & $\cdots$ & $\cdots$ & 103/2 & 1/2 & 103/16 &
 {\bf I} \\
$3640$ & $5356$ & $4608942$ & $880588800$ & $\cdots$ & $\cdots$ & 52 & 1/2 & 13/2  &
 {\bf I} \\
\hline
\hline
$3675$ & $5460$ & $4791255$ & $910793625$ & $\cdots$ & $\cdots$ & 105/2 & 1/2 & 105/16 & 
 {\bf I} \\
$3710$ & $5565$ & $4978926$ & $941672200$ &$\cdots$ & $\cdots$ &  53 & 1/2 & 53/8  &
 {\bf I} \\
\hline
\hline
$3745$ & $5671$ & $5172059$ & $973231875$ & $\cdots$ & $\cdots$ & 107/2 & 1/2 & 107/16 &
 {\bf I} \\
$3780$ & $5778$ & $5370759$ & $1005480000$ & $\cdots$ & $\cdots$ & 54 & 1/2 & 27/4 & {\bf I} \\
\hline
\hline
$3815$ & $5886$ & $5575132$ & $1038423925$ & $\cdots$ & $\cdots$ &  109/2 & 1/2 & 109/16 &  {\bf I} \\
 $3850$ & $5995$ & $5785285$ & $1072071000$ & $\cdots$ & $\cdots$ & 55 & 1/2 & 55/8  & 
 {\bf I} \\
 \hline
 \hline
3864 &  13110 & 13725282 & 2591631168 & \tiny{12971091} & \tiny{4897835680923668} & 276/5 & 4/5 & 33/5 & {\bf III} \\
$3885$ & $6105$ & $6001326$ & $1106428575$ & $\cdots$ & $\cdots$ &  111/2 & 1/2 & 111/16 &  {\bf I} \\
 $3920$ & $6216$ & $6223364$ & $1141504000$ &  $\cdots$ & $\cdots$ & 56 & 1/2 & 7  &
 {\bf I} \\
\hline
\hline
\end{tabular} 
\end{threeparttable}
\end{center}
\end{table}

\begin{table}[h] 
\begin{center}
\begin{threeparttable}
\caption{Solutions to the $\mathbf{[3,2]}$ MLDE with $\mathbf{0<c\leq 96}$}\label{t15}
\begin{tabular*}{\textwidth}{@{\extracolsep{\fill}}cccc||cc||ccc||l}
\hline
\hline
\makebox[0pt][l]{\fboxsep0pt\colorbox{Mywhite} {\strut\hspace*{\linewidth}}}
$N$ &  $m_1$ & $m_2$ & $k$ & $D_1$ & $D_2$ & $c$ & $h_1$ & $h_2$ & CFT  \\
\hline
\hline
\makebox[0pt][l]{\fboxsep0pt\colorbox{Mywhite} {\strut\hspace*{\linewidth}}}
$84$ & $1$ & $1$ & $737352$ & $\cdots$ & $\cdots$ & 2/5 & 1/60 & 1/5 & ${ \bf \mathcal{M}(5, 2)} \oplus {\rm constant}$ \\
\makebox[0pt][l]{\fboxsep0pt\colorbox{Mygrey} {\strut\hspace*{\linewidth}}}
$210$ & $3$ & $4$ & $2058000$ & $\cdots$ & $\cdots$ & $1$ & 1/24 & 1/4 &  $(\mathbf{\hat{A}_{1}})_1 \oplus {\rm constant}$ \\
\makebox[0pt][l]{\fboxsep0pt\colorbox{Mywhite} {\strut\hspace*{\linewidth}}}
$420$ & $8$ & $17$ &$4762800$ & $\cdots$ & $\cdots$ & $2$ & $1/12$ & ${1}/{3}$ & $(\mathbf{\hat{A}_{2}})_1 \oplus {\rm constant}$ \\
\makebox[0pt][l]{\fboxsep0pt\colorbox{Mygrey} {\strut\hspace*{\linewidth}}}
$588$ & $14$ & $42$ & $7277088$ & $\cdots$ & $\cdots$ & ${14}/{5}$ & ${7}/{60}$ & ${2}/{5}$ & $(\mathbf{\hat{G}_{2}})_1 \oplus {\rm constant}$ \\
\makebox[0pt][l]{\fboxsep0pt\colorbox{Mywhite} {\strut\hspace*{\linewidth}}}
$840$ & $28$ & $134$ & $11289600$ & $\cdots$ & $\cdots$ & $4$ & ${1}/{6}$ & ${1}/{2}$ & $(\mathbf{\hat{D}_{4}})_1 \oplus {\rm constant}$ \\
\makebox[0pt][l]{\fboxsep0pt\colorbox{Mygrey} {\strut\hspace*{\linewidth}}}
$1092$ & $52$ & $377$ & $14768208$ & $\cdots$ & $\cdots$ & ${26}/{5} $ & ${13}/{60}$ & ${3}/{5}$ &  $(\mathbf{\hat{F}_{4}})_1 \oplus {\rm constant}$ \\
\makebox[0pt][l]{\fboxsep0pt\colorbox{Mywhite} {\strut\hspace*{\linewidth}}}
$1260$ & $78$ & $729$ & $15876000$  & $\cdots$ & $\cdots$ & $6$ & ${1}/{4}$ & ${2}/{3}$ & $(\mathbf{\hat{E}_{6}})_1 \oplus {\rm constant}$  \\
\makebox[0pt][l]{\fboxsep0pt\colorbox{Mygrey} {\strut\hspace*{\linewidth}}}
$1470$ & $133$ & $1673$ & $13582800$ & $\cdots$ & $\cdots$ & $7$ & ${7}/{24}$ & ${3}/{4}$ &  $(\mathbf{\hat{E}_{7}})_1 \oplus {\rm constant}$ \\
\makebox[0pt][l]{\fboxsep0pt\colorbox{Mywhite} {\strut\hspace*{\linewidth}}}
$1596$ & $190$ & $2831$ & $7775712$ & $\cdots$ & $\cdots$ & ${38}/{5}$ & ${19}/{60}$ & ${4}/{5}$ &$\mathbf{E_{7\frac12}} \oplus {\rm constant}$  \\
\hline
\hline
\end{tabular*} 
\end{threeparttable}
\end{center}
\end{table}

\end{document}